\documentclass[11pt]{article}

\usepackage{amsmath,amssymb}

\newcommand{\R}{\mathbb{R}}
\newcommand{\C}{\mathbb{C}}

\newcommand{\A}{{\cal A}}
\newcommand{\E}{{\cal E}}
\newcommand{\Abar}{\overline{\cal A}}

\newcommand{\h}{{\cal H}}
\newcommand{\scal}[2]{\langle #1| #2\rangle}

\newcommand{\cyl}{{\rm Cyl}}
\newcommand{\Cyl}{{\rm Cyl}}

\newcommand{\scripta}{\mathfrak{A}}

\newcounter{mnotecount}[section]

\newtheorem{lm}{Lemma}
\newtheorem{df}{Definition}
\newtheorem{thr}{Theorem}
\newtheorem{cor}{Corollary}
\newtheorem{chr}{Characterization}

\def\tE{\tilde{E}}
\def\tphi{\tilde{\varphi}}
\def\be#1\ee{\begin{equation}#1\end{equation}}

\numberwithin{equation}{section}
\numberwithin{lm}{section}
\numberwithin{thr}{section}
\numberwithin{chr}{section}
\numberwithin{df}{section}

\begin{document}

\title{Diffeomorphism covariant representations of the holonomy-flux
$\star$-algebra}
\author{ Andrzej Oko{\l}\'ow${}^{1*}$
 and Jerzy\ Lewandowski${}^{2,1,3\dagger}$}
\date{February 14, 2003}

\maketitle
\begin{center}
{\it 1. Instytut  Fizyki Teoretycznej, Uniwersytet
Warszawski, ul. Ho\.{z}a 69, 00-681 Warsaw, Poland\\
2. Center for Gravitational Physics and Geometry, Physics
Department, 104 Davey, Penn State, University
Park, PA 16802, USA\\
3. Max--Planck--Institut f\"ur Gravitationsphysik, Albert--Einstein--Institut,
 14476 Golm, Germany\\
${}^*$ oko@fuw.edu.pl\\
${}^\dagger$ lewand@fuw.edu.pl}
\end{center}
\medskip

\begin{abstract} Recently, Sahlmann \cite{sahl-1} proposed a new,
algebraic point of view on the loop quantization. He brought up
the issue of a $\star$-algebra underlying that framework, studied
the algebra consisting of the fluxes and holonomies and
characterized its representations.  We  define the diffeomorphism
covariance of a representation of the Sahlmann algebra and study
the diffeomorphism covariant representations. We prove they are
all given by Sahlmann's decomposition into the cyclic
representations of the sub-algebra of the holonomies by using  a
single state only. The state corresponds to the natural measure
defined on the space of the generalized  connections. This result
is a generalization of Sahlmann's result \cite{sahl-2} concerning
the $U(1)$ case.
\end{abstract}


\section{Introduction}


The quantum holonomy operators and the quantum flux operators are
the basic elements of the quantum geometry
\cite{al-difgeom,geomgen,area,vol,acz} and loop quantum
gravity \cite{lqg}, the theory also  called the Quantum Spin
Dynamics (QSD) after the title of Thiemann's series  of works
\cite{QSD}.
 Quantum geometry is
a quantum theory of  geometry of initial data of
Einstein's theory in terms of the Ashtekar variables. The flux
quantum operators correspond to the intrinsic geometry
(to the  fluxes of the classical triads)
 whereas the
holonomy operators to the extrinsic curvature (the Riemannian
connection is combined with the extrinsic curvature into a new
$SU(2)$-connection 1-form). In the standard representation, the
operators are defined  in the Hilbert space  given by  the natural
diffeomorfism invariant  measure denoted in this paper by
$\mu_{\rm AL}$ \cite{al-meas} defined on the Ashtekar-Isham
 quantum configuration space of generalized
connections  \cite{ai,al-meas,al-difgeom}. Every holonomy operator is just the
multiplication by the corresponding holonomy variable. Each flux
operator is  $i$ times certain derivation defined on the
differentiable functions of the holonomies \cite{area}. These flux 
operators are self-adjoint. All the basic operators of the quantum
geometry (volume, area, length),  the scalar constraint as well as
the quantum Hamiltonian operators of matter coupled with 
quantum geometry \cite{QSD} are constructed by using those elementary
holonomy-flux operators \cite{area,vol}. The framework
can be also generalized to a loop quantization of diffeomorphism
invariant theories of $G$-connections defined over a
$d$-dimensional manifold \cite{lqg}. Recently, Sahlmann \cite{sahl-1}
proposed a new, more algebraic point of view.  It opens the door
to a representation theory of the quantum geometry. The main idea
is to spell out a definition of a $\star$-algebra constructed from
the holonomies and fluxes, that  underlies all the loop quantum
gravity framework, and to study its representations. The
conclusion is that, upon some quite natural assumptions, every
representation of the Sahlmann algebra can be characterized by
using the direct sum of cyclic representations of the
Ashtekar-Isham  commutative $C^\star$-algebra of the holonomies.
Each cyclic representation is defined in a Hilbert space given by
a measure on the space of the generalized connections.  The main
result of Sahlmann is establishing  a general form in which the
flux operators can be represented. Each flux operator is given by
the corresponding derivation {\it  plus}  a suitable
multiplication operator (we often call it a correction term).
Finally, in the  case of the structure group $U(1)$ of the theory,
Sahlmann proved that every diffeomorphism covariant representation
necessarily corresponds to the natural measure mentioned above
\cite{sahl-2}.

In our paper we introduce a precise definition the  holonomy-flux
$\star$-algebra  in the way following from \cite{sahl-1}. We name
it the Sahlmann algebra. According to the definition, the
commutative subalgebra of the holonomies contained in the Sahlmann
algebra is not the entire Ashtekar-Isham $C^\star$-algebra.
Therefore a priori, the representation theory of commutative
$C^\star$-algebras does not apply. We fill that gap, and show that
every representation of the subalgebra extends uniquely to a
representation of the entire $C^\star$-algebra of the holonomies.
In this way me merge Sahlmann's characterization of the
holonomy-flux $\star$-algebra. Up to now, diffeomorphism invariance and covariance are not well defined notions in the context of representations. To see why this is so, first observe that the action of (analytic) diffeomorphisms in the Sahlmann algebra induces an action on the representations. We do not expect a representation to be invariant, but it is not even clear how to define covariance of a representation. This is because diffeomorphisms do not act naturally in the carrier space of a given representation (Sahlmann's characterization depends on a choice of the cyclic vectors).  We propose a solution of that issue and
formulate a definition of a representation of the Sahlmann algebra
covariant with  respect to the diffeomorphisms. On a working
level, our definition coincides with the one used by Sahlmann in
the $U(1)$ case mentioned above. Our main result is a proof that
every diffeomorphism covariant representation of the holonomy-flux
$\star$-algebra, upon  an assumption about the domain, is given by
Sahlmann's characterization such that all the measures used in the
direct sum are the same natural measure $\mu_{\rm AL}$. This
result is valid in the case of the covariance with respect to the
group of the analytic diffeomorphisms of $\Sigma=\R^d$ ($d\geq
2$), and for an arbitrary compact, connected Lie group. The only
remaining ambiguity is in the definition of the representation of
the flux operators in the correction term. However, the
identification of the carrier Hilbert space as an orthogonal
product of the spaces given by the natural measure $\mu_{\rm AL}$
enables one to use the powerful machinery of the decompositions
into the spin-network functions. It can be used to reduce
considerably the remaining freedom in the representation of the
quantum fluxes. One more separate result of our paper is that the
only cyclic representation of the holonomy-flux $\star$-algebra
$\scripta$ given by the identity map is the one given by the
natural measure (see Lemma \ref{lm-al}).

The motivation for our work came from numerous discussions
with Ha\-n\-no Sahlmann about his published and unpublished
results. For an independent account on the diffeomorphism
covariant representations  of the holonomy-flux-algebra see
\cite{sahlthiem}.


\section{Preliminaries}

The loop quantization framework was introduced and developed in
\cite{al-difgeom}-\cite{ai}. The goal of the program is  the canonical
quantization of diffeomorphism invariant  and gauge invariant
theories of connections. The main application is loop quantum
gravity. Then, in the $3+1$ case, the phase space is
the cotangent bundle of the space  of $SU(2)$-connections defined
on a (trivial) bundle over a $3$-dimensional manifold. In this
paper we deal with the arbitrarily dimensional case. For the
simplicity we consider $G$-connections, where $G$ is a compact,
connected Lie group, defined on a trivial bundle over a
$d$-dimensional manifold $\Sigma$. The generalization to a
non-trivial bundle is also possible \cite{B,al-difgeom}. The framework
considered in this paper applies to any theory, whose phase space
$\A\times \E$ consists of pairs of canonically conjugate fields
$(A,\tilde{E})$  defined on  $\Sigma$, where: $A$ is a  Lie
algebra $G'$ valued differential $1$-form  and $\tilde{E}$ is a
vector density of weight $1$ which takes values in the  space
$G'{}^*$ dual to the Lie algebra $G'$. That is, the Poisson
bracket between two  functions $H$ and $L$ defined on $\A\times
\E$ is given by the standard relation:
\begin{equation}
\{H, L \}\ =\ \int_\Sigma dx\left(\frac{\delta H}{\delta A^i_a(x)}
\frac{\delta L}{\delta \tE^a_i(x)}-\frac{\delta L}{\delta A^i_a(x)}
\frac{\delta H}{\delta \tE^a_i(x)}\right),
\label{can}
\end{equation}
where the fields $A$ and $\tilde{E}$ are expressed by their components with
respect to a basis $(\tau_i)$, $i=1,...,n$ of the Lie algebra $G'$, the dual
to $(\tau_i)$ basis $(\tau^{*i})$ of $G'{}^*$, and a (local) coordinate chart
$(x^a)$, $a=1,...,d$ on $\Sigma$, i.e.:
\[
A=A^i_a\tau_i\otimes dx^a\;\;\;\tilde{E}=\tilde{E}^a_i\tau^{*i}\otimes
\partial_a.
\]
The integrand is a density of the weight 1, hence the integral
does not depend on choice of a coordinate chard $(x^a)$.
Every map $a:\Sigma\rightarrow G$ defines a gauge transformation:
\[
(A,\tE)\ \mapsto\ (a^{-1}Aa+ a^{-1}da, a^{-1}\tE a)
\]
where, being more formal, given $v\in G'$, the element  $a^{-1}va$
is obtained by the adjoint action ${\rm ad}(a^{-1})v$ of $G$ in
$G'$, and the $G'$ valued differential 1-form $a^{-1}da$ is the
pullback $a^*\theta_{{\rm MC}}$ of the Maurer-Cartan form
$\theta_{{\rm MC}}$. Those transformations are often called ``the
Yang--Mills'' gauge transformations, to emphasize that the  group
of all the gauge transformations of the theory (in the sense of
the hamiltonian  approach) is bigger and  contains also the
diffeomorphisms.

The fields $A\in\A$ and $\tilde{E}\in\E$ are referred to as
connections and, respectively, electric fields. In the loop
quantization the space  $\A$  of connections is chosen to be a
configuration space and $\E$ becomes the  space of momenta. In
this paper we do not consider any other fields coupled to
$(A,\tE)$, however, in the presence of other fields the framework presented
still applies to the $(A,\tE)$ degrees of freedom and
there is known a natural extension of this approach which
accommodates the coupled fields \cite{QSD}.


\subsection{The holonomy-flux variables}


The first step in the loop quantization is a non-standard choice
of elementary
 classical variables that will have direct quantum analogs. They are the
functions of finitely many holonomies along curves in $\Sigma$---the
cylindrical functions---and fluxes of the electric field across $(d-1)$-dimensional
submanifolds in
 $\Sigma$. We recall now the definitions in detail.

Consider an oriented 1-dimensional submanifold $e$ with boundary given by an embedding of an interval: $e:[t_0,t_1]\rightarrow \Sigma$. It is called an {\it edge}. The holonomy transport $h_{e}(t,t_0,A)$ along an the edge $e$ from $e(t_0)$ to $e(t_1)$ with respect to a connection $A$ is an element of $G$ defined by the following differential equation and initial  conditon:
\begin{subequations}
\begin{align}
\frac{d}{dt}h_{e}(t,t_0,A)\ &=\ -A_a(e(t))\dot{e}^a(t)h_{A,e}(t,t_0),\\
                 h_e(t_0,t_0,A)\ &=\ \mathbb{I}
\end{align}
\end{subequations}
where $\mathbb{I}\in G$ stands for the neutral element of $G$. Given $A\in\A$, the parallel transport
along an edge $e$ is denoted by:
\begin{equation}
A(e)\ :=\ h_e(t_1,t_0,A)={\cal P}\exp (-\int_e A),
\label{hol}
\end{equation}
and called {\it the holonomy variable}. The holonomy variable $A(e)$ is insensitive on  orientation preserving reparametrizations of $e$. Denote the splitting of an edge $e$ into two edges $e_1$, $e_2$ of the same orientation  by $e=e_2\circ e_1$, and the edge  obtained by reversing the orientation of $e$ by $e^{-1}$. The holonomy variable satisfies the following relations:
\[
A(e_2)A(e_1)\ =\ A(e_2\circ e_1), \ \ A(e^{-1})\ =\ A(e)^{-1}.
\]
\begin{df}
A complex valued function $\Psi$ defined on $\A$ is called {\it cylindrical}, if there is a finite set of edges $\{e_1,...,e_N\}$ and a function $\psi$ defined on $G^N$ such that:
\be
\Psi(A)\ =\ \psi(A(e_1),...,A(e_N)).
\label{cyl}
\ee
We say $\Psi$ is compatible with the set of edges $\{e_1,...,e_N\}$.
\end{df}

We consider in this work the set $\cyl^\infty$ of all the cylindrical
functions given by all the $C^\infty$ functions $\psi$ in (\ref{cyl})
(analogously, we define the set $\cyl^0$ of all the cylindrical functions
given by all
the continuous functions $\psi$) . $\cyl^\infty$ is a subalgebra of the
$\star$-algebra of all the complex-valued  functions defined on $\A$ (with
$\star$ being the complex conjugation). A naturally defined norm in
$\cyl^\infty$ is the sup norm,
\[
\|\psi\|_{\rm sup}\ :=\ \sup_{A\in\A}|\Psi(A)|.
\]
The completion $\overline{\cyl^\infty}$ in that norm together with $\star$
and the norm is the Ashtekar-Isham   $C^\star$-algebra denoted by $\cyl$ below.
The Gel'fand-Naimark spectrum of that algebra is characterized
as the space ${\Abar}$ of the generalized connections being homomorphisms between
the groupoid of the curves in $\Sigma$ and  the gauge group $G$
\cite{al-difgeom} (see also Section \ref{gen-conn}).

To define the {\it flux of} $E$ fix: $(i)$ a finite $(d-1)$ submanifold $S\subset \Sigma$  $(ii)$ an orientation in the normal bundle to $S$, and $(iii)$ a function $f:S\rightarrow G'$.

The corresponding flux of $\tE$ is defined to be\footnote{ $\epsilon_{a a_1 \ldots a_{d-1}}$ is a Levi-Civita density of weight $-1$ on $\Sigma$.}
\[
E(S,f)\ :=\ \frac{1}{(d-1)!}\int_S\tilde{E}_i^{a}f^i\epsilon_{a a_1 \ldots a_{d-1}}dx^{a_{1}}\wedge\ldots\wedge dx^{a_{d-1}},
\]
where the (local) orientation of $S$ used in the integral is determined
by the orientation of the normal bundle and the orientation of a given local
coordinate chart.


\subsection{The Ashtekar-Corichi-Zapata Poisson algebra}


Now we turn to the Poisson bracket relations between the
elementary classical variables. It should be emphasized, however,
that the fluxes and the cylindrical functions are in fact
distributional objects from the point of view of the Poisson
bracket (\ref{can}). Therefore, to  apply the Poisson bracket
(\ref{can}) to the fluxes and holonomies some extension procedure
is needed. Ashtekar, Corichi and Zapata \cite{acz} proposed a very
natural passage. They use the observation that  every function
defined on the phase space $\A\times \E$ and linear in
$\tilde{E}$ defines a derivation $X$ in the algebra of functions
defined on $\A$, namely \be \{\Phi,P_X\}\ =:\ X\Phi, \label{1} \ee
where the function linear in $\tilde{E}$ is denoted by $P_X$.
Then, for
 functions $\Phi,\, \Phi'$ independent of $\tilde{E}$,  and any two functions
 $P_X,\, P_{X'}$ linear in $\tilde{E}$, the following relations are true,
\begin{equation}
\{P_X,P_{X'}\}\ =\ P_{[X,X']}, \ \ \{\Phi, \Phi'\}\ =\ 0.
\label{2}
\end{equation}
Therefore, assigning  to each of the linear functions $P_X$ the
derivation $X$, and to each of the $\tE$ independent functions
$\Phi$ the multiplication operator by the given function $\Phi$
defines an isomorphism between: $a)$ the Poisson algebra spanned
by the functions $P_X$ and $\Phi$,  and $b)$ the Lie algebra of
operators spanned by the derivations $X$, and the multiplication
by the functions $\Phi$ operators. The isomorphism maps $\{,\}
\mapsto [,] $. The idea of \cite{acz} is to define an algebra of
the classical elementary variables from the cylindrical functions
and the flux variables as a Lie algebra  of operators analogous to
the one described above. Indeed, as we explain below, every
electric flux does define through the Poisson bracket (\ref{1}) a
non-trivial derivation operator in the algebra $\Cyl^\infty$. We
start from the discussion of that operator.

For the Poisson bracket (\ref{1}) between a cylindrical function
$\Psi$ compatible with a set of edges $\{e_1,...,e_N\}$ and the
flux  $E(S,f)$ across a submanifold $S$  to be finite, the
intersection between the edges $\{e_1,...,e_N\}$ and $S$ has to
satisfy some regularity property. To ensure that property
arbitrarily, we {\it fix an analytic structure on $\Sigma$} and
define  {\it regular analytic submanifolds} in the following way:
$(i)$  {\it $0$-dimensional regular analytic sumbanifold of $\Sigma$ is a one-element subset of $\Sigma$ $(ii)$ an $m$-dimensional ($0<m<\dim\Sigma$) regular analytic submanifold $M$ is an analytic orientable $m$-submanifold  such that its closure $\overline{M}$ is a compact subset of an analytic $m$-dimensional submanifold of $\Sigma$ and $\overline{M}\setminus M$ is a finite sum of regular analytic submanifolds of the lower dimensions.}
We assume that all the edges used in
the definition of the  cylindrical functions
are the closures of the regular analytic 1-submanifolds.
We also assume that the family of all the submanifolds $S$ used in the flux definition contains only   
regular analytic $(d-1)$-submanifolds (later, this family will be extended
to a larger one by set differences and intersections of the regular analytic $S$'s).   

Upon the regular analyticity assumption, every cylindrical
function on $\A$ is compatible with an analytically embedded
graph, whose definition we recall below.

\begin{df}
An analytically embedded graph $\gamma$ in $\Sigma$ is a set of
analytic edges $\{e_1,\ldots,e_N\}$ in $\Sigma$ such that two
distinct edges can meet each other only in their endpoints. The
edges  will be called the edges of the graph $\gamma$, and their
endpoints the vertices of $\gamma$.
\end{df}

To calculate the Poisson bracket between a given cylindrical
function $\Psi\in\cyl^\infty$ and the flux function $E(S,f)$
across a surface $S$, we select a graph compatible with $\Psi$
appropriately. Given $\Psi$ and $S$ we choose a graph $\gamma$
compatible with $\Psi$ such that every edge of $\gamma$ either:
$(i)$ is contained  in $S$ (modulo its end points), or $(ii)$
intersects $S$ at exactly one  endpoint, or $(iii)$ does not
intersect $\gamma$ at all. The orientation in the normal bundle
$S$ is used to distinguish the subset of the class $(ii)$ edges
placed `up' (`down') the surface $S$. Those edges will be labelled
by $I_{\rm up}$ ( $I_{\rm down}$). After the substitution of
$\Psi$ and $E(S,f)$ for the functions $H$ and $L$ in (\ref{can}) the
direct calculation gives the following result,
\begin{multline}
\{\Psi, E(S,f)\}= \frac{1}{2}\sum_{v\in S}f^i(v)
\left(\sum_{e_{I_{\rm down}}}X_{i,v,e_{I_{\rm down}}} -
\sum_{e_{I_{\rm up}}}X_{i,v,e_{I_{\rm up}}}\right)
\Psi =:\\
=: X_{S,f}\Psi
\label{X}
\end{multline}
where given $v$, $e_{I_{\rm up/down}}$ runs  through the subset of
the edges of $\gamma$ such that $v$ is an isolated intersection
between $e_{I_{\rm up/down}}$ and $S$, and
\begin{multline*}
\left(X_{i,v,e_I}\Psi\right)(A(e_1),\ldots,A(e_I),\ldots,A(e_N))\ :=\\
:=
\begin{cases}
\frac{d}{ds}\big|_{s=0}\,\Psi(\ldots,A(e_I)\exp(s\tau_i),\ldots),&
\text{if $v$ is the beginning of $e_I$}\\
\frac{d}{ds}\big|_{s=0}\,\Psi(\ldots,\exp(-s\tau_i)A(e_I),\ldots),&
\text{if $v$ is the end of $e_I$}\end{cases}
\end{multline*}

In conclusion, each electric field flux $E(S,f)$ corresponds to
the derivation  $X_{S,f}: \cyl^\infty\rightarrow \cyl^\infty$.
With this result we go back to the construction of the algebra of
the classical elementary observables from the cylindrical
functions and fluxes.

Consider the complex vector space spanned by  the operators:
\begin{gather*}
\Cyl^\infty\ni \Psi'\mapsto \Psi \Psi'\in \Cyl^\infty,\\
\Cyl^\infty\ni \Psi'\mapsto X_{S,f} \Psi'\in\Cyl^\infty
\end{gather*}
given by all the cylindrical functions $\Psi\in \Cyl^\infty$ and
all the fluxes $E(S,f)$ corresponding to all submanifolds $S$ and
smearing functions $f$. Extend the resulting space of operators to
the (smallest possible) operator Lie algebra, and denote it by
$\scripta_{\rm ACZ}$.

\begin{df}
The  algebra $(\scripta_{\rm ACZ},\{,\}=[,])$ is the
Ashtekar-Corichi-Za\-pa\-ta  algebra of the
elementary classical variables whose  classical bracket $\{,\}$ is
by definition the commutator (no $i$ factor thus far) between the
derivation and/or multiplication operators.
\end{df}

 The
Ashtekar-Corichi-Zapata  algebra can be written as the direct sum
\[
\scripta_{\rm ACZ}\ =\ \cyl^\infty\oplus {\cal X},
\]
where ${\cal X}$ are  the derivations obtained by taking all the
$X_{S,f}$ operators given by all the submanifolds $S$ and functions
$f$ and all the (multiple) commutators.

The derivation operator
$X_{S,f}$ was defined for a $(d-1)$-dimensional {\em submanifold} $S$.
However, the definition (\ref{X}) of the operator 
is naturally extended to $(d-1)$-dimensional  {\em manifold with boundary} embedded in $\Sigma$. Indeed,  consider two
operators $X_{S,f}$ and  $X_{S',f'}$ such that $S'\subset S$
and $f_{|S'}=f'$. Then the operator
\be
X_{S\setminus S',f}\ :=\ X_{S,f}\ -\ X_{S',f'}
\ee
is an operator of a flux across $S\setminus S'$ being an embedded manifold with boundary.

 A (classical)
anomaly in the Ashtekar-Corichi-Zapata algebra is that a
single classical elementary variable may correspond to two
different derivations. An example is a surface $S$ split  by an
edge $e$ into two pieces, $S=S_1\cup e\cup S_2$. Then,
$E(S,f)=E(S_1,f)+S(S_2,f)$, but $X_{S,f}\not=
X_{S_1,f}+X_{S_2,f}$. Another kind of anomaly are the commutators
$[X_{S_1,f_1},X_{S_2,f_2}]$ when $S_1\cap S_2$ is 1-dimensional.

\section{The Sahlmann holonomy-flux $\star$-algebra}

The quantization consists in assigning to every elementary
classical variable, that is to every element of the
Ashtekar-Corichi-Zapata  Lie algebra $(\scripta_{\rm ACZ},
\{,\})$, an element of an appropriate $\star$-algebra ($\scripta,
\star$),
\[
\hat{{}}\ :\ \scripta_{\rm ACZ}\ \rightarrow \scripta,
\]
such that
\[
\widehat{\{a,b\}}\ =\ i[\hat{a},\hat{b}].
\]

In this section we give  a precise definition of the $\star$-algebra
$\scripta$  following (modulo a small improvement) the
definition proposed by Sahlmann  \cite{sahl-1} after whom we name
the algebra. Sahlmann's results of \cite{sahl-1}
will provide a useful  characterization of  representations
of $\scripta$. However, before we conclude that, we  need to
show that every representation of the algebra
$\cyl^\infty$ admits a decomposition into the cyclic
representations as every representation of the full commutative
$C^\star$-algebra $\Cyl$. Next, we outline Sahlmann's characterization
of representations of $\scripta$.


\subsection{Definitions}\label{def}


Consider the subalgebra $\scripta$ of the algebra of the linear
operators $\cyl^\infty\rightarrow\cyl^\infty$ generated by the
Ashtekar-Corichi-Zapata algebra. Introduce the $\star$-operator in
$\scripta$ by specifying its action on the generating set:
\[
\Psi^\star\ :=\ \bar{\Psi}, \ \ X_{S,f}^\star\ =\ -X_{S,f},
\]
where $\Psi\in \cyl^\infty$ and $X_{S,f}$ is the derivative
operator defined in (\ref{X}). The $\star$ operation extends consistently
and uniquely to an involutive anti-isomorphism of the entire
algebra $\scripta$.
\begin{df}\label{def-sah}
The Sahlmann holonomy-flux $\star$ algebra is the algebra $(\scripta, \star)$.
The quantization map  $\ \hat{{}}:\scripta_{\rm ACZ}\rightarrow \scripta$ is
defined on the generators of the Ashtekar-Corichi-Zapata algebra
as follows:
\begin{equation}
\hat{\Psi}\ :=\ \Psi,\ \ \hat{X}_{S,f}\ :=\ -iX_{S,f},
\label{hat}
\end{equation}
\end{df}
This definition concludes an important part of the quantization.
The classical elementary variables are assigned to appropriate quantum
elementary variables:
\begin{equation}
\begin{cases}
\Psi\mapsto\hat{\Psi}&\\
E(S,f)\mapsto\hat{X}_{S,f}&
\end{cases}.
\label{class-op}
\end{equation}
Throughout this paper, all the elements of the Sahlmann holonomy-flux
$\star$-algebra will be denoted by hatted letters $\hat{a},\hat{b},...$,
in the way consistent with (\ref{hat}).

The last step of the kinematical loop quantization is construction
of a $*$-representation of the holonomy-flux $\star$-algebra.  Because the Sahlmann algebra is not equipped with any norm we are not allowed to expect that one can represent the algebra by means of {\em bounded} operators on a Hilbert space. This leads to the following definition of a $*$-representation of $\scripta$:
\begin{df}
Let $L(\h)$ be a space of linear operators on a Hilbert space $\h$. We say that a map $\pi:\scripta\rightarrow L(\h)$ is a $*$-representation of $\scripta$ on the Hilbert space $\h$ if: 
\begin{enumerate}
\item there exists a dense subspace ${\cal D}$ of $\h$ such that   
\[
{\cal D}\subset\bigcap_{\hat{a}\in\scripta}[\ D(\pi(\hat{a}))\cap D(\pi^*(\hat{a}))\ ], 
\]
where $D(\pi(\hat{a}))$ denotes the domain of the operator $\pi(\hat{a})$;

\item for every $\hat{a},\hat{b}\in\scripta$ and $\lambda\in\C$ the following conditions are satisfied on ${\cal D}$:
\begin{alignat*}{3}
\pi(\hat{a}+\hat{b})&=\pi(\hat{a})+\pi(\hat{b}),&\qquad\pi(\lambda\hat{a})&=\lambda\pi(\hat{a}),\\
\pi(\hat{a}\hat{b})&=\pi(\hat{a})\pi(\hat{b}),&\qquad \pi(\hat{a}^\star)&=\pi^*(\hat{a}).
\end{alignat*}  
\item If given $\hat{a}\in\scripta$ there exists a dense linear subspace $D\subset D(\pi(\hat{a}))$ such that $\pi(\hat{a})|_D$ is closable, then $\pi(\hat{a})$ is equal to the closure of $\pi(\hat{a})|_D$.     
\end{enumerate}
We will say that $\pi$ is nondegenerate iff $\pi(\hat{a})v=0$ for every $\hat{a}\in\scripta$ implies $v=0$.  
\label{repr-df}
\end{df}  
These conditions mean in particular, that $\pi(\hat{a}){\cal D}\subset{\cal D}$ and that every element $\hat{a}=\hat{a}^\star$ is represented by the {\em symmetric} operator $\pi(\hat{a})$.
 
In the sequel we will consider only nondegenerate $*$-representations of $\scripta$.

There is known a natural (nondegenerate) $*$-representation $\pi_{\rm AL}$ defined by the natural
measure $\mu_{\rm AL}$ introduced \cite{area,vol}  in the
Gel'fand-Neimark spectrum ${\Abar}$ of $\Cyl$ (see also Section
\ref{gen-conn}). The domain of the representation is
$\Cyl^\infty\subset L^2({\Abar},\mu_{\rm AL})$ and $\pi_{\rm AL}$
is just the identity map. The representation has a large group of
symmetries containing all the analytic diffeomorphisms of $\Sigma$
(we devote Section \ref{diff-sym} to the issue of the
diffemorphism invariance/covariance). Remarkably, the natural measure
$\mu_{\rm AL}$ is the only measure with respect to which the
identity map {\it is} a $\star$-algebra representation. We prove
this uniqueness in Section \ref{which}.


\subsection{Representations of $\Cyl^\infty$}


Recently, Sahlmann started a new program of  systematic study of
the representation theory of the holonomy-flux $\star$-algebra
$\scripta$. To merge Sahlmann's
conclusions on the representation theory of the $\star$-algebra
$\scripta$ we need one connecting lemma (this is a consequence of
the difference in definitions between the current algebra
$\scripta$ and that of \cite{sahl-1}). 

Notice first that if $\pi$ is a nondegenerate $*$-representation of $\scripta$ and $v\in{\cal D}$ then for every $\hat{a}$ 
\[
\pi(\hat{a})(v-\pi(\hat{1})v)=0
\]    
($\hat{1}$ is the unit of the Sahlmann algebra defined by the function $\cyl^\infty\ni\Psi\equiv 1$), hence by virtue of the nondegeneracy assumption $\pi(\hat{1})={\rm id}$ on $\cal D$ and consequently on the whole $\h$. This means in particular that the $*$-representation $\varrho$ of $\cyl^\infty$ defined as $\varrho:=\pi|_{\cyl^\infty}$ is nondegenerate.    

\begin{lm}
Suppose $\varrho$ is a nondegenerate $*$-representation of $\cyl^\infty$ on $\h$ in the sense of Definition \ref{repr-df}. Then $\varrho$ maps $\cyl^\infty$ into the $C^\star$-algebra $B(\h)$ of the bounded operators on $\h$ and admits a unique extension to a $C^\star$-algebra homomorphism of $\Cyl$ into $B(\h)$. 
\label{pi-infty}
\end{lm}

\medskip
\noindent{\bf Proof.} We will show that for every $\Psi\in \Cyl^\infty$
and  every  $v\in {\cal D}$ the following inequality is true,
\be \|\varrho(\Psi)v\|_{\h}\ \le\
\|\Psi\|_{\rm sup}\,\|v\|_{\h},
\label{pi}
\ee
where our notation distinguishes the Hilbert space norm in $\h$
from the sup-norm in the $C^\star$-algebra $\Cyl$. Given the inequality,
it is obvious that: $(i)$ for every $\Psi\in\Cyl^\infty$ the
operator $\varrho(\Psi)$ admits a unique extension to a bounded
operator defined in $\h$,  $(ii)$ $\varrho$ admits a unique extension
to a linear and continuous map $\varrho:\Cyl\rightarrow B(\h)$, and
$(iii)$ it is straightforward to check that the resulting
extension of $\varrho$ onto $\Cyl$ is a $C^\star$-algebra
homomorphism.

To show (\ref{pi}) fix an arbitrary number $q>1$ and notice that
for every $\Psi\in\Cyl^\infty$ there exists an element $\Phi\in
\Cyl^\infty$ such that\footnote{This fact was pointed out to us by
Professor S. L. Woronowicz.}
\be\label{trick} \Phi\Phi^*\ +\ \Psi\Psi^*\ =\ q^2
\,\|\Psi\|^2_{\rm sup}.
 \ee
Indeed, according to the definition of $\Cyl^\infty$, there is a
graph $\gamma=\{e_1,...,e_N\}$  and a smooth function
$\psi:G^N\rightarrow \C$, such that
$\Psi(A)=\psi(A(e_1),...,A(e_N))$ for every $A\in\A$.
Define a new function $\Phi(A)=\phi(A(e_1),...,A(e_N))$
where
\be \phi \ :=\  \sqrt{q^2\, \|\psi \|_{\rm sup}^2\ -\
|\psi |^2}. \ee
Notice that the function under the square root is non-negative
and never vanishes, therefore the square is uniquely defined and
the smoothness of $\psi$ implies the smoothness
$\phi$.

Using (\ref{trick}) and the equality $\varrho(1)={\rm id}$ we find that for arbitrary $v\in {\cal D}$:
\be \|\varrho(\Psi)v\|^2_{\h}\ =\ q^2\|\Psi\|_{\rm sup}^2\,\|v\|^2_{\h} -
\|\varrho(\Phi)v\|^2_{\h}\ \le q^2\|\Psi\|_{\rm sup}^2\,\|v\|^2_{\h}. \ee
Given $\Psi$ and $v$, the inequality holds for every $q>1$,
therefore it is true also for $q=1$. This completes the proof of
the lemma. $\blacksquare$


\subsection{Sahlmann's characterization of representations of $\scripta$}


Starting from this point we can directly apply Sahlmann's
arguments \cite{sahl-1} leading to a useful characterization of a quite
broad class of representations of the holonomy-flux $\star$-algebra
$\scripta$. Suppose $\pi$ is a nondegenerate $\ast$-representation of
$\scripta$ in a Hilbert space $\h$. Consider first
the restriction of $\pi$ to the subalgebra $\Cyl^\infty$. According to Lemma \ref{pi-infty},
$\pi|_{\cyl^\infty}$ extends to a representation of the $C^\star$-algebra
$\cyl$ in $\h$, therefore it admits the following decomposition described by
\begin{chr}[Sahlmann]\mbox{}\smallskip
\begin{enumerate}
\item Representation $\pi|_{\cyl^\infty}$ is the direct sum of cyclic representations
$\{\pi_\nu\}$:
\be\label{A}
\h=\bigoplus_{\nu\in{\cal N}}\h_\nu,\;\;\;\pi|_{\cyl^\infty}=
\bigoplus_{\nu\in {\cal N}}\pi_\nu,
\ee
where $\{\h_\nu\}$ are carrier spaces of representations $\{\pi_\nu\}$
 respectively,  $\nu$ ranges some label set ${\cal N}$ and the sum
is orthogonal;
\item For each $\nu$:
\be\label{B}
\h_\nu\  = \  L^2(\Abar,\mu_\nu),
\ee
where $\Abar$ is a Gel'fand-Neimark spectrum\footnote{For more details
about the space $\Abar$ see Subsection \ref{gen-conn}.} of $\cyl$, and
$\mu_\nu$ is a regular, Borel measure on $\Abar$ defined by a positive, linear functional
 $\cyl\rightarrow \C$.
\item Let $\Psi\in L^2(\Abar,\mu_\nu)$ and $\Phi\in\cyl^\infty$. Then:
\be\label{C}
\pi(\hat{\Phi})\Psi=\Phi\Psi.
\ee
\end{enumerate}
\label{dec}
\end{chr}

Given $\Psi\in \h$ we will denote by $\Psi_\nu$ the orthogonal projection
of $\Psi$ onto $\h_\nu$, and write $\Psi=(\Psi_\nu)$.
\medskip

Turn now to the whole representation $\pi$. 
Following \cite{sahl-1} we define:
\begin{multline}
{\cal C}^\infty:=\{ (\Psi_\nu)\in \h \ \big|\; \text{$\Psi_\nu=0$ for all but finitely many $\nu$'s}\\ \text{and $\Psi_\nu\in \cyl^\infty$ for every $\nu$}\}.
\label{c-infty}
\end{multline}
It turns out that the assumption ${\cal D}={\cal C}^\infty$, where $\cal D$ is introduced by Definition \ref{repr-df}, gives the following useful characterization:  

\begin{chr}[Sahlmann]
Suppose $\pi$ is a nondegenerate $*$-rep\-re\-sen\-ta\-tion  of $\scripta$ in a Hilbert
space $\h$, and $\pi$ satisfies the assumption ${\cal D}={\cal C}^\infty$. Then $\h$
and the restriction of $\pi$ to $\Cyl^\infty$ satisfy the
decomposition given by Characterization \ref{dec}. Moreover, for every $\hat{X}_{S,f}$,
$\pi$ defines a family of elements of $\h$ labelled by elements of
the labelling set ${\cal N}$ (the same as in (\ref{A},\ref{B})) \be {\cal
N}\ni\iota\mapsto {F_{S,f}}^\iota\in \h, \ee such that the
following conditions are satisfied (given ${F_{S,f}}^\iota$  we
will subsequently denote by ${{F_{S,f}}^\iota}_\nu$ the $\h_\nu$
component of ${F_{S,f}}^\iota$):

\begin{enumerate}
\item for every $\Psi=(\Psi_\nu)\in{\cal C}^\infty$:
\[
\pi(\hat{X}_{S,f})\Psi=\hat{\mathbf{X}}_{S,f}\Psi+\hat{F}_{S,f}\Psi
\]
where $\hat{\mathbf{X}}_{S,f}\Psi:=(\hat{X}_{S,f}\Psi_\nu)$ and:
\[
\hat{F}_{S,f}\Psi=\hat{F}_{S,f}(\Psi_\nu):=
(\sum_\iota\Psi_{\iota}{F_{S,f}}^{\iota}\!_{\nu})
\]
where ${F_{S,f}}^{\iota}\!_{\nu}\in L^2(\Abar,\mu_\nu)$;
\item for every $\Phi,\Phi'\in \cyl^\infty\subset L^2(\Abar,\mu_\nu)$:
\begin{equation}
\scal{\hat{X}_{S,f}\Phi}{\Phi'}_\nu-\scal{\Phi}{\hat{X}_{S,f}\Phi'}_\nu=
\scal{\Phi}{({F_{S,f}}^\nu\!_\nu-{\overline{F}_{S,f}}^\nu\!_\nu)\Phi'}_\nu,
\label{div-X}
\end{equation}
where $\scal{\cdot}{\cdot}_\nu$ is the scalar product on
$L^2(\Abar,\mu_\nu)$ ({\bf \em no} summation with respect to the
index $\nu\in{\cal N}$ in ${F_{S,f}}^\nu\!_\nu$);
\item for every $S=S_1\cup S_2$ such that  $S_i$ ($i=1,2$) are disjoint:
\begin{equation}
{{F_{S,f}}^{\iota}}_\nu={{F_{S_1,f_1}}^{\iota}}_\nu+{{F_{S_2,f_2}}^{\iota}}_\nu,
\label{F-sum}
\end{equation}
where $f_i:=f|_{S_i}$.
\end{enumerate}
\label{sahl-th}
\end{chr}

Sahlmann's characterization of \cite{sahl-1} contains a more
exhaustive discussion; we only outlined those elements which will
be relevant in the current paper.

From the point of view of the loop quantization program, fixing
a specific representation $\pi$ of the Sahlmann holonomy-flux $\star$-algebra
completes the (kinematical) quantization. The composition of $\pi$ with
the mapping $\ \hat{{}}\ $ (\ref{class-op}) becomes  a quantum representation
of the classical elementary variables as quantum operators in $\h$:
\begin{equation}
\begin{cases}
\Phi\mapsto\pi(\hat{\Phi})&\\
E_{S,f}\mapsto\pi(\hat{X}_{S,f}) &
\end{cases}.
\label{repr-class}
\end{equation}
In particular,  $\pi(\hat{X}_{S,f})$ is the {\em quantum flux operator}.


\section{A diffeomorphism symmetry and statement of the problem \label{diff-sym}}



\subsection{Statement of the problem}

The smooth diffeomorphisms of $\Sigma$ act naturally in the space
of connections $\A$. The action induces an action of the analytic
diffeomorphisms in the $C^\star$-algebra $\Cyl$ and finally in the
Sahlmann holonomy-flux $\star$-algebra $\scripta$.( In this paper we consider
the analytic diffeomorphisms only and henceforth we will drop
the word `analytic'.)  However, imposing the diffeomorphism
covariance on a representation $\pi$ of the Sahlmann algebra is a
priori ambiguous. Therefore, we formulate now a precise definition
of the diffeomorphism covariance considered in this paper. Given a
diffeomorphism $\varphi:\Sigma\rightarrow\Sigma$, denote by
$\tphi$ the induced action $\tphi:\cyl^\infty\rightarrow
\cyl^\infty$. Consider first a simple case when
$\pi_{|\Cyl^\infty}$ is a cyclic representation, that is when the
decomposition (\ref{A},\ref{B}) consists of a single term only.
The representation $\pi$ will be called diffeomorphism covariant
if there is a cyclic element $v\in \h$ such that the state it
defines in $\scripta$ is diffeomorphism invariant:
\[
\scal{v}{\pi(\tphi\hat{a}\tphi^{-1})v}\ =\ \scal{v}{\pi(\hat{a})v}
\]
for every $\hat{a}\in \scripta$ and for every diffeomorphism $\varphi$.
A general
definition we formulate is:

\begin{df}
Suppose $\pi$ is a  nondegenerate $*$-representation of the Sahl\-mann
$(\scripta,\star)$ in a Hilbert space $\h$ described by
Characterization \ref{dec}. The representation $\pi$ is called covariant
with respect to the diffeomorphisms of $\Sigma$ if each of the components $\h_\nu$ in (\ref{A},\ref{B}) contains a cyclic\footnote{We mean here, that $v_\nu$ is cyclic with respect to the corresponding representation $\pi_\nu$ defined by (\ref{A}).} vector
$v_\nu$ such that every finite linear combination $v$ of the
vectors $v_\nu$, $\nu\in {\cal N}$ defines a diffeomorphism
invariant state in $\scripta$; that is, when:
\[
\scal{v}{\pi(\tphi\hat{a}\tphi^{-1})v}\ =\
\scal{v}{\pi(\hat{a})v}
\]
for every  $a\in \scripta$, and every diffeomorphism $\varphi$.
\label{df-inv}
\end{df}

We will see in Subsection \ref{work} that every diffeomorphism
maps a diffeomorphism covariant representation into
a unitarily equivalent one. In this sense the covariance implies
the invariance.

Importantly, the representation used in the loop quantization program---which is given by $\h=L^2(\Abar,\mu_{\rm AL})$ with  the natural measure
$\mu_{\rm AL}$ and all the correction functions $F_{S,f}{}^\iota{}_\nu=0$ ---{\it is} diffeomorphism covariant in the sense of Definition \ref{df-inv}.

A working version of the
definition of the covariance was assumed by Sahl\-mann in \cite{sahl-2}.
He shows that in the $G=U(1)$ case a diffeomorhism covariant
cyclic representation $\pi$ necessarily corresponds to the natural
measure $\mu_{\rm AL}$ mentioned in the Section \ref{def} (see
also the next subsection).

The goal  of this paper is to study the diffeomorphism covariant
representations $\pi$ in the general case of  Characterization \ref{sahl-th}
and to derive the resulting restrictions on the measures $\mu_\nu$
in the decomposition (\ref{A},\ref{B}) and on the corrections
$F_{S,f}{}^\iota{}_\nu\in L^2({\Abar},\mu_\nu)$ to the action of
the flux operators in the general compact and connected gauge
group $G$ case.

In the remaining part of this section we outline the
facts concerning  the measures on ${\Abar}$ and  the space ${\Abar}$ itself
which will be used in the next section. In particular we recall
the definition of the natural measure $\mu_{\rm AL}$. Finally, we
formulate the diffeomorpfism invariance/covariance
condition in terms of  Characterization \ref{sahl-th}.


\subsection{The measures on  $\Abar$ \label{gen-conn}}

The $C^\star$-algebra $\cyl$ is isomorphic with the algebra of the
continuous functions on the compact Hausdorff space $\Abar$ of the
generalized connections. The space $\Abar$ can be naturally
identified with the space of maps $\bar{A}: {\bf E} \rightarrow
G$, where ${\bf E}$ is the space of the edges in $\Sigma$, such
that \cite{al-difgeom}:
\begin{gather*}
\bar{A}(e^{-1})=[\bar{A}(e)]^{-1}\\
\bar{A}(e_2\circ e_1)=\bar{A}(e_2)\bar{A}(e_1)
\end{gather*}
for every triple of edges $e_1,\,e_2,\,e$. Obviously, an example
is the holonomy map (\ref{hol}). Given $\bar{A}$, the
corresponding element of $\Abar$ is the homomorphism
$\Cyl^0\ni\Psi\mapsto\psi(\bar{A}(e_1),...,\bar{A}(e_N))\in \C$ where $\Psi$ is
written as in (\ref{cyl}).

A convenient tool for description measures on $\Abar$ is a family
of quotient spaces labelled by the graphs in $\Sigma$, obtained
from $\Abar$ and equipped with a family of some projective maps. A
graph  $\gamma =\{e_1,\ldots,e_N\}$ defines an equivalence
relation on $\Abar$:
\[
\bar{A}_1\sim_\gamma\bar{A}_2\Longleftrightarrow \bar{A}_1(e_I)=\bar{A}_2(e_I)
\]
for every $I=1,\ldots,N$. This relation defines the quotient space
$\Abar_\gamma$,
\[
\Abar_\gamma:=\Abar/\sim_\gamma,
\]
and the projection $p_\gamma:{\Abar}\rightarrow \Abar_\gamma$.
The map $\Abar\ni \bar{A}\mapsto (\bar{A}(e_1),...,\bar{A}(e_N))$
induces a bijective map
\begin{equation}
\Abar_\gamma\leftrightarrow G^N,
\label{G}
\end{equation}
which equips the space $\Abar_\gamma$ with the geometry of $G^N$.
Moreover, for every equivalence class $[\bar{A}]_\gamma\in\Abar_\gamma$
there exists $A\in\A$ such that $A\in[\bar{A}]_\gamma$ \cite{al-meas,B}.
Thus every function $\psi:G^N\rightarrow\C$ defining cylindrical function
$\Psi$ compatible with a graph $\gamma$ can be regarded as a function on
$\Abar_\gamma$ (see Definition \ref{cyl}).

Finally, we recall the projective family structure of the family
of the spaces $\{\Abar_\gamma\}$ labelled by all the graphs
$\gamma$ in $\Sigma$.  Notice first that graphs in $\Sigma$ form
a partially ordered, directed set with relation $\geq$ defined as
follows: $\gamma'\geq \gamma$ if every edge of $\gamma$ can be
expressed as a composition of edges of $\gamma'$ (and their
inverses) and each vertex of $\gamma$ is a vertex of $\gamma'$.
Now, if $\gamma'\geq\gamma$ then there exists a projection
$p_{\gamma\gamma'}:\Abar_{\gamma'}\rightarrow\Abar_\gamma$ such
that:
\begin{equation}
p_\gamma= p_{\gamma\gamma'}\circ p_{\gamma'}.
\label{proj-map}
\end{equation}

An important application of
$\{\Abar_\gamma,p_\gamma,p_{\gamma\gamma'}\}$ is the construction of
measures on $\Abar$ from projective families of measures defined
on the spaces $\Abar_\gamma$. Every measure $\mu$ defined on
$\Abar$ defines a family of measures  $\{\mu_\gamma:=p_{\gamma\ast}\mu\}$
on $\{\Abar_\gamma\}$ respectively:
\[
\int_{\Abar_\gamma}\psi\;d\mu_\gamma =\int_{\Abar}\Psi\;d\mu,
\]
for every $\Psi\in\cyl^0$ compatible with a graph $\gamma$ (in the latter equation
$\Psi$ is related to $\psi$  as in Definition \ref{cyl}). If $\Psi$ is compatible
with two graphs $\gamma,\gamma'$, then:
\begin{equation}
\int_{\Abar_{\gamma}}\psi\;d\mu_{\gamma} =\int_{\Abar_{\gamma'}}\psi'\;d\mu_{\gamma'}.
\label{two-gra}
\end{equation}
Thus in the case of $\gamma'\geq\gamma$ we have:
\be(p_{\gamma\gamma'})_*\mu_{\gamma'}=\mu_\gamma \label{cons-fam}.
\ee

Conversely, if for every graph $\gamma$ we endow the space
$\Abar_\gamma$ with a measure $\mu_\gamma$, such that the
resulting family of measures is consistent in the sense of Equation (\ref{cons-fam})
for every pair of graphs $\gamma'\ge\gamma$, then the following integration functional:

\be \Cyl^0\ni\Psi\mapsto \int_{\Abar_\gamma}\psi\;
d\mu_\gamma \in \C
\ee
is positive definite hence defines a regular, Borel measure on $\Abar$
\cite{al-meas}.

The natural measure $\mu_{\rm AL}$ is defined by choosing each
$\mu_\gamma$ to be the measure given by the probability Haar
measure on $G^N$ and by the bijection (\ref{G}).


\subsection{The action of the diffeomorphisms in the Sahlmann
algebra}


Let $\varphi$ be a diffeomorphism on $\Sigma$. It acts in a
natural way on pair of fields $(A,\tilde{E})$:
\begin{equation}
A\mapsto\varphi^\ast A,\;\;\;\tilde{E}\mapsto\varphi^{-1}_\ast \tE.
\label{diff-var}
\end{equation}
This induces an action $\tilde{\varphi}$ of $\varphi$ on the
cylindrical functions (\ref{cyl}) and the fluxes:
\begin{gather}
(\tilde{\varphi}\Phi)(A):=
\phi(A(\varphi(e_1)),...,A(\varphi(e_N)))\label{diff-fcyl}\\
\tilde{\varphi}E(S,f)\ :=\
E(\tilde{S},\tilde{f})\nonumber,\\
\tilde{S}=\varphi(S),\;\;\;\tilde{f}=(\varphi^{-1}|_S)^\ast f.\nonumber
\end{gather}
The derivations $X_{S,f}$ are
defined in a diffeomorphism covariant way; that is, for every
$\Phi\in\cyl^\infty$, flux $E(S,f)$ and  diffeomorphism $\varphi$
we have
\begin{equation}
\tilde{\varphi}(X_{S,f}\Phi)=X_{\tilde{S},\tilde{f}}\tilde{\varphi}\Phi.
\label{varphi-X}
\end{equation}

Notice that by using the characterization of the space $\Abar$ as
the space of maps $\bar{A}: {\bf E} \rightarrow G$, the
diffemorphisms of $\Sigma$ act naturally in  $\Abar$ via
$(\varphi^*\bar{A})(e):=\bar{A}(\varphi(e))$. This action is
consistent with the extension of the cylindrical functions to
functions defined on $\Abar$ and with the action
(\ref{diff-fcyl}).

The Sahlmann algebra $\scripta$ is a subalgebra of linear operators
acting in $\cyl^\infty$, thus the action
$\tphi:\cyl^\infty\rightarrow\cyl^\infty$  induces the following
action of $\varphi$ on $\scripta$:
\[
\scripta\ni\hat{a}\mapsto\tphi\hat{a}\tphi^{-1}\in\scripta.
\]
The action is a clearly a $\star$-algebra isomorphism. In
particular, we have the following action for the generators. For every $\hat{\Phi}$, where
$\Phi\in\cyl^\infty$, and for every derivation operator
$\hat{X}_{S,f}$, the action reads
\begin{equation}
\tphi\hat{\Phi}\tphi^{-1}\ =\ \widehat{\tphi\Phi},\ \
\tphi\hat{X}_{S,f}\tphi^{-1}\ =\ \hat{X}_{\tilde{S},\tilde{f}}.
\label{diff-hPhihX}
\end{equation}


\subsection{The diffeomorphism covariance  conditions in
terms of Characterization \ref{dec} \label{work}}


Consider now a representation $\pi$ of the Sahlmann
$\star$-algebra $\scripta$ in a Hilbert space
$\h\bigoplus_{\nu\in{\cal N}}\h_\nu$ given by Characterization
\ref{dec}
and assume it is covariant with respect to the diffeomorphisms of
$\Sigma$ in the sense of Definition \ref{df-inv}. The family of
the  vectors $v_\nu$, $\nu\in {\cal N}$ provided by the
definition can be used to induce a unitary action $U_\varphi$ of a
diffeomorphism  $\varphi$ of $\Sigma$  in $\h$. Then, the
conditions of Definition \ref{df-inv} turn into appropriate
invariance/covariance conditions with respect that action. We
spell them out in this subsection.

Notice first that we have the following identification (see (\ref{B})):
\begin{equation}
L^2(\Abar,\mu_\nu)\supset\cyl\ni\Phi \ \leftrightarrow \  \pi(\hat{\Phi})v_\nu\in\h_\nu.
\label{iso}
\end{equation} 
Using this it is easy to see that Definition \ref{df-inv} implies the equality ${\cal D}={\cal C}^\infty$. Consequently Characterization \ref{sahl-th} is true as well.	 

The induced action $u_\varphi$
of a diffeomorphism $\varphi$ in each  $\h_\nu$  is defined by:
\[
u_\varphi\pi(\hat{\Phi})v_\nu:=
\pi(\tphi\hat{\Phi}\tphi^{-1})v_\nu=\pi(\widehat{\tphi\Phi})v_\nu,
\]
where $\Phi\in\cyl$ and the existence of a unique extension of $\pi_{|\Cyl^\infty}$
to the entire $\Cyl^\star$ algebra completion  $\Cyl$ is used. The
corresponding action of $\varphi$ in $\h$ will be denoted by
$U_\varphi$.

The first consequence of the diffeomorphism covariance
of the representation $\pi$ (Definition \ref{df-inv}) is that,
$\varphi\mapsto U_\varphi$ is a unitary
representation of the diffeomorphism group in $\h$, and
$\varphi\mapsto u_\varphi$ is a unitary representation of the
diffeomorphism group in $\h_\nu$ for each cyclic component
$\h_\nu$.

It is easy to see that the second consequence of the
diffeomorphism covariance  is that $\pi$ is
covariant with the actions of the diffeomorphisms in the algebra
$\scripta$ and in the Hilbert space $\h$. That is, for every
diffeomorphism $\varphi$  and every $\hat{a}\in \scripta$ the equality
\begin{equation}
\pi(\tphi\hat{a}\tphi^{-1})=U_\varphi\pi(\hat{a})U_\varphi^{-1}
\label{inv7}
\end{equation}
holds on ${\cal C}^\infty$.
 
Every diffeomorphism invariant state $v_\nu$ defines a diffeomorphism
invariant measure on $\Abar$:
\be
\int_{\Abar} \Psi d\mu_\nu\ :=\ \scal{v_\nu}{\pi(\hat{\Psi})v_\nu}.
\ee
Indeed,  in terms of the isomorphism (\ref{iso})
the action of  $u_\varphi$ in $\h_\nu$   coincides with the action
of $\tphi$ in $L^2(\Abar,\mu_\nu)$. The unitarity of $u_\varphi$ is equivalent to the invariance of the
measures $\mu_\nu$, $\nu\in {\cal N}$ with respect to the action
$\tphi$ of each diffeomorphism $\varphi$.

Now we can state that:

\begin{cor}
If $\pi$ is a nondegenerate, diffeomorphism covariant $*$-rep\-re\-sen\-ta\-tion of $\scripta$, then $\pi$ satisfies requirements of Characterizations \ref{dec} and \ref{sahl-th}. Moreover, the three conditions: $(i)$ the unitarity of the operators $U_\varphi$, $(ii)$ ${\cal D}={\cal C}^\infty$ and $(iii)$ the diffeomorphism
covariance condition $(\ref{inv7})$ are equivalent to  Definition \ref{df-inv}. 
\label{cor}
\end{cor}

In particular, taking into account (\ref{diff-hPhihX}) we have for every operator $\hat{X}_{S,f}$,
\begin{equation}
\pi(\hat{X}_{\tilde{S},\tilde{f}})=U_\varphi\pi(\hat{X}_{S,f})U^{-1}_\varphi.
\label{u-esf}
\end{equation}
(the latter and the next two equations hold on ${\cal C}^\infty$). On the other hand (\ref{varphi-X}) implies:
\[
\hat{\mathbf{X}}_{\tilde{S},\tilde{f}}=U_\varphi\hat{\mathbf{X}}_{S,f}U^{-1}_\varphi.
\]
Since  $\pi(\hat{X}_{S,f})=\hat{\mathbf{X}}_{S,f}+\hat{F}_{S,f}$ according to
Characterization \ref{sahl-th}, the diffeomorphism covariance
of $\pi$ implies the following covariance of
$\hat{F}_{S,f}$:
\begin{equation}
\hat{F}_{\tilde{S},\tilde{f}}=U_\varphi\hat{F}_{S,f}U^{-1}_\varphi,
\label{diff-hF}
\end{equation}
where $\tilde{S}=\varphi(S),\,\tilde{f}=(\varphi^{-1}|_S)^\ast f$.


\section{Main theorem}


Let us formulate the theorem which is the main result of this paper.

We consider the Sahlmann $\star$-algebra $(\scripta,\,\star)$
(Definition \ref{def-sah}) corresponding to the space
of the   $G$-connections defined on the trivial bundle $\Sigma\times G$
where $G$ is  a compact, connected Lie group. The natural measure
defined on the space  $\Abar$ of the generalized connections was recalled in
Section \ref{gen-conn}.

\begin{thr}
Suppose $\pi$ is a nondegenerate $*$-representation of the  Sahl\-mann  algebra
$\scripta$. Suppose also, that
$\pi$ is covariant with respect to the group of the (analytic)
diffeomorphisms of $\Sigma$ in the meaning of Definition \ref{df-inv},
and $\Sigma=\R^d$. Then, for every $\nu\in {\cal N}$, the measure $\mu_\nu$
in Characterization \ref{dec}  corresponding to the diffeomorphism
invariant state $v_\nu$ of Definition \ref{df-inv} is  the natural measure:
\[
\mu_\nu\ =\ \mu_{\rm AL}
\]
defined on $\Abar$. In the consequence, all the $L^2(\Abar,\mu_{\rm AL})$
functions ${{F_{S,f}}^\nu}_\nu$ (no summation) used in Characterization
\ref{sahl-th} are real valued.
\label{main}
\end{thr}

\noindent{\bf Remarks}
\begin{enumerate}
\item The only reason for the (restrictive) assumption $\Sigma=\R^d$ is
that in the proof below, we will need
some special {\em analytic} diffeomorphisms on $\R^d$.
Thus far we were able to construct the needed diffeomorphisms
only in the $\Sigma=\R^d$ case.
\item Taking into account Lemma \ref{pi-infty}  we can see that upon 
the requirement of nondegeneracy and the  defintion of the diffeomorphism covariance, every representation $\pi$ of the Sahlmann $\star$-algebra satisfies the conclusions of the Theorem \ref{main}. 
\end{enumerate}


\section{Proof of the main theorem}


Corollary \ref{cor} guarantees that every representation $\pi$ satisfying assumptions of the main theorem satisfies conditions of Characterizations \ref{dec} and \ref{sahl-th}. 
According to the latter one the quantum flux operator $\pi(\hat{X}_{S,f})$ has the following form:
\begin{equation}
\pi(\hat{X}_{S,f})=\hat{\mathbf{X}}_{S,f}+\hat{F}_{S,f}
\label{sahl-res}
\end{equation}
for every submanifold $S$ and smearing function $f$. 

Our proof consists of two steps.
First,  we will show that whenever $S$ is diffeomorphic
to a $(d-1)$-dimensional (coordinate) cube $C$ and $f:S\rightarrow G'$
is a constant function, then the imaginary part of ${F_{C,f}}^\nu\!_\nu$ has
to be equal zero for every $\nu$. Therefore,
the corresponding operator $\hat{X}_{C,f}$ restricted to $\cyl^\infty \subset L^2(\Abar,\mu_\nu)$ 
is symmetric with respect to the scalar product on $L^2(\Abar,\mu_\nu)$.

The symmetry of the operator $\hat{X}_{C,f}$ will be enough to conclude
in the second step of the proof that, for every $\nu$, the measure
$\mu_\nu$ is the natural measure $\mu_{\rm AL}$.


\subsection{Step 1: The imaginary part of ${F_{C,f}}^\nu\!_\nu$}



\subsubsection{The functions ${F_{S,f}}^\iota\!_\nu$}


Consider the functions  ${F_{S,f}}^\iota\!_\nu$ and their imaginary parts:
\[
{I_{S,f}}^\iota\!_\nu:=\frac{1}{2}({F_{S,f}}^\iota\!_\nu-{\overline{F}_{S,f}}^\iota\!_\nu)
\in L^2(\Abar,\mu_\nu).
\]

The assumed diffeomorphism covariance of representation $\pi$ allows us
to make use of the results derived in Subsection \ref{work}. In particular Equation (\ref{diff-hF}) implies
\[
U_\varphi\hat{F}_{S,f}\Psi=\hat{F}_{\tilde{S},\tilde{f}}U_\varphi\Psi,
\]
for every diffeomorphism $\varphi$ and for every $\Psi\in{\cal C}^\infty$. The l.h.s. of the above equation can be transformed in the following way:
\begin{multline*}
U_\varphi\hat{F}_{S,f}\Psi=
U_\varphi(\sum_\iota{F_{S,f}}^\iota\!_\nu\Psi_\iota)=
(\sum_\iota u_\varphi({F_{S,f}}^\iota\!_\nu\Psi_\iota))=\\=(\sum_\iota
u_\varphi({F_{S,f}}^\iota\!_\nu) u_\varphi(\Psi_\iota)),
\end{multline*}
while the r.h.s. is:
\[
\hat{F}_{\tilde{S},\tilde{f}}U_\varphi\Psi=\hat{F}_{\tilde{S},\tilde{f}}
U_\varphi(\Psi_\nu)=\hat{F}_{\tilde{S},\tilde{f}}(u_\varphi\Psi_\nu)=(\sum_\iota
{F_{\tilde{S},\tilde{f}}}^\iota\!_\nu u_\varphi(\Psi_\iota)).
\]
A comparison of the two latter results shows that the
elements ${F_{S,f}}^\iota$ of  $\h$ are assigned
to the submanifolds $S$ and the functions $f$ in a covariant way,
\begin{equation}
u_\varphi{F_{S,f}}^\iota\!_\nu={F_{\tilde{S},\tilde{f}}}^\iota\!_\nu, \ \
{\rm hence}\ \
u_\varphi{I_{S,f}}^\iota\!_\nu={I_{\tilde{S},\tilde{f}}}^\iota\!_\nu.
\label{I-cov}
\end{equation}
where $\tilde{S}=\varphi(S)$ and $\tilde{f}=(\varphi^{-1}|_S)^\ast f$.

The diffemorphism covariance of the representation $\pi$ also implies
that for every $\nu$
the scalar product $\scal{\cdot}{\cdot}_\nu$ on $L^2(\Abar,\mu_\nu)$ is
diffeomorphism
invariant, thus
\begin{equation}
||{I_{S,f}}^\iota\!_\nu||_\nu=||u_\varphi{I_{S,f}}^\iota\!_\nu||_\nu=||{I_{\tilde{S},\tilde{f}}}^\iota\!_\nu||_\nu.
\label{square}
\end{equation}

Equation (\ref{F-sum}) allows us to conclude, that for $S=S_1\cup S_2$, where $S_1$
and $S_2$ are  disjoint
\begin{equation}
{I_{S,f}}^\iota\!_\nu={I_{S_1,f_1}}^\iota\!_\nu + {I_{S_2,f_2}}^\iota\!_\nu
\label{sum-I}
\end{equation}
(here $f_i:=f|_{S_i}$ and $i=1,2$).

Using the properties (\ref{square},\,\ref{sum-I}) we will show in the next
subsections, that for every $(d-1)$-dimensional cube $C$ in $\R^d$ and every constant function $f$
\[
{I_{C,f}}^\nu\!_\nu=0.
\]


\subsubsection{The functions ${I_{C,f}}^\nu\!_\nu$ for cubes  and
constant functions $f$\label{cubes}}


Consider in $\Sigma=\R^d$ an arbitrary coordinate system
 $(x^j)$, ($j=1,\ldots,d$)
given by an affine transformation
applied to the Cartesian coordinates.
Let  $C\subset \R^d$ be a cube given  by the following inequalities:
\be\label{cube}
C:=\{y\in \R^d\;\big|\;
\begin{cases}
x^d(y)=0&\\
-l< x^1(y) \leq l &\\
-l< x^i(y) < l & \text{for $i=2,3,\ldots,d-1$.}
\end{cases};
\;\;\;l>0\}.
\ee

Notice, that $C$ includes one of the sites, namely the one
contained in the $x^d=0, x^1=l$ plane. Due to that subtlety,
$C$ can be split by a suitable partition into  two disjoint cubes
$C_1,C_2$ diffeomorphic to $C$. The cubes can be defined as follows:
\begin{gather*}
C_1:=\{y\in C \; \big| \; x^1(y)\leq 0\}\\
C_2:=\{y\in C \; \big| \; x^1 (y)> 0\}.
\end{gather*}

Let us fix the index $\nu$ and denote $I_{C,f}={I_{C,f}}^\nu\!_\nu$ in order to simplify
notation. Recall that $I_{C,f}\in L^2(\Abar,\mu_\nu)$. We are going now to find some
 relation between  $I_{C,f}$ and $I_{C_i,f|_{C_i}}$ ($i=1,2$) for a constant function $f$.

Assume then that $f$ is an arbitrary constant function on $C$.  Because
the cubes $C_1,C_2$
are diffeomorphic to $C$, then Equation (\ref{square}) gives us:
\[
||I_{C,f}||^2_\nu=||I_{C_1,f}||^2_\nu=||I_{C_2,f}||^2_\nu,
\]
$C=C_1\cup C_2$ and $C_1\cap C_2=\varnothing$, thus by virtue of Equation (\ref{sum-I}):
\[
I_{C,f}=I_{C_1,f}+I_{C_2,f}.
\]
Combining the two latter equations we obtain:
\[
||I_{C,f}||^2_\nu=||I_{C_1,f}+I_{C_2,f}||^2_\nu=2||I_{C,f}||^2_\nu+\scal{I_{C_1,f}}
{I_{C_2,f}}_\nu+\scal{I_{C_2,f}}{I_{C_1,f}}_\nu,
\]
that is:
\begin{equation}
||I_{C,f}||^2_\nu=-2\scal{I_{C_1,f}}{I_{C_2,f}}_\nu.
\label{square-I}
\end{equation}

Our goal now is to show, that the scalar product $\scal{I_{C_1,f}}{I_{C_2,f}}_\nu$ is equal 0. We will do it in two steps. First, we will express the function $I_{C_1,f}$ as a limit of a suitably chosen sequence of cylindrical functions belonging to
$\Cyl^\infty\subset L^2(\Abar,\mu_\nu)$. Then, Characterization
\ref{sahl-th} will allow us to conclude, that
$\scal{I_{C_1,f}}{I_{C_2,f}}_\nu=0$.


\subsubsection{The function $I_{C_1,f}$ as a limit of a sequence of
cylindrical functions}


The function $I_{C_1,f}$ can be expressed as a limit,
\[
I_{C_1,f}=\lim_{n\rightarrow\infty}\Phi_n;\;\;\;\Phi_n\in\cyl^{\infty}.
\]
Notice that in fact we have quite a large freedom in a choice of the
sequence converging to $I_{C_1,f}$; to see this consider a
sequence of (analytic) diffeomorphisms
$(\varphi_n):\R^d\rightarrow\R^d$  such that every $\varphi_n$
preserves the cube $C_1$. Then, applying  (\ref{I-cov}) and
the unitarity of $u_\varphi$ we obtain:
\[
||I_{C_1,f}-\Phi_n||_\nu=||u_{\varphi_n}(I_{C_1,f}-\Phi_n)||_\nu=||I_{C_1,f}-u_{\varphi_n}(\Phi_n)||_\nu,
\]
which means that $u_{\varphi_n}(\Phi_n)$ converges to $I_{C_1,f}$
as well. We will use this freedom to construct a special
sequence which converges to $I_{C_1,f}$.

Let us fix a number $n$ and consider a graph $\gamma_n$ compatible with
the cylindrical function $\Phi_n$. In general, some edges of the graph can be
transversal to the cube $C_2$, i.e. they can have isolated intersection
points with $C_2$. Then the action of $\hat{X}_{C_2,f}$ on $\Phi_n$ is
nontrivial,
\[
\hat{X}_{C_2,f}\Phi_n\neq 0.
\]

\begin{lm}
For each of the graphs $\gamma_n$, $n=1,...,$ defined above,
there exists an analytic
diffeomorphism $\varphi_n:\R^d\rightarrow\R^d$,
such that:
\begin{enumerate}
\item $\varphi_n$ preserves the cube $C_1$;
\item the graph  $\varphi_n(\gamma_n)$ has no edges transversal to the cube $C_2$.
\end{enumerate}
\label{diff-lm}
\end{lm}
Such a diffeomorphism is constructed explicitly in Appendix.

Use now the diffeomorphisms $(\varphi_n)$ given by Lemma \ref{diff-lm}
to construct the following  sequence convergent to $I_{C_1,f}$,
\[
\tilde{\Phi}_n:=u_{\varphi_n}(\Phi_n).
\]
Now, each function in the sequence $(\tilde{\Phi}_{n})$ is compatible with
a graph having no edge transversal to the cube $C_2$, hence:
\begin{equation}
\hat{X}_{C_2,f}\tilde{\Phi}_n=0.
\label{Xphi}
\end{equation}


\subsubsection{The vanishing of
$\scal{I_{C_1,f}}{I_{C_2,f}}_\nu$\label{scal-sec}}


Equation (\ref{div-X}) in the case  $\Phi'=1$ gives
\[
\scal{\hat{X}_{C_2,f}\Phi}{1}_\nu=2\scal{\Phi}{I_{C_2,f}}_\nu,
\]
for every smooth cylindrical function $\Phi\in L^2(\Abar,\mu_\nu)$. Owing to
Equation (\ref{Xphi})
\[
0=\lim_{n\rightarrow\infty}\scal{\hat{X}_{C_2,f}
\tilde{\Phi}_n}{1}_\nu=\lim_{n\rightarrow\infty}2\scal{\tilde{\Phi}_n}
{I_{C_2,f}}_\nu=2\scal{I_{C_1,f}}{I_{C_2,f}}_\nu.
\]
This result together with Equation (\ref{square-I}) imply
\begin{equation}
||I_{C,f}||^2_\nu=0, \ \ {\rm hence}\ \   I_{C,f}={I_{C,f}}^\nu\!_\nu=0,
\label{I-zero}
\end{equation}
(where the last equality refers to elements of $L^2(\Abar,\mu_\nu)$
while the measure $\mu_\nu$ is not assumed to be faithful),
and completes the first  step of  the proof.


\subsection{Step 2: Which measures on $\Abar$ admit a symmetric action of
$\hat{X}_{C,f}$? \label{which}}

Equation (\ref{div-X}) and (\ref{I-zero}) imply, that for any constant
function $f$ and
for every smooth cylindrical functions $\Phi,\Phi'$:
\[
\scal{\hat{X}_{C,f}\Phi}{\Phi'}_\nu-\scal{\Phi}{\hat{X}_{C,f}\Phi'}_\nu=0.
\]
This  means that for every surface $C'$ diffeomorphic to
the cube (\ref{cube})
the operator $\hat{X}_{C',f}$ is symmetric on $\cyl^\infty\subset
L^2(\Abar,\mu_\nu)$.
To complete  the proof of (or disprove) Theorem \ref{main} we have to answer the
question
asked in the
title of this subsection. The answer is given by the  lemma
we formulate now,
where by a cube, we mean every surface $C'\in\R^d$ which
is diffeomorphic to the cube $C$ considered in Section \ref{cubes}.

\begin{lm}
Let $G$ be a compact, connected group. Suppose that, for every
$(d-1)$-dimensional cube $C$ in $\R^d$ and for every constant function
$f:C\rightarrow G'$ the  operator $\hat{X}_{C,f}$
is symmetric on $\cyl^\infty\subset L^2(\Abar,\mu)$.  Then
\[
\mu=\mu_{\rm AL}.
\]
\label{lm-al}
\end{lm}

The lemma means, that for every $\nu\in {\cal N}$ the measure
$\mu_\nu=\mu_{\rm AL}$, so the proof of the lemma completes the
proof of the first conclusion of the main theorem.
 On the other hand, for {\it every} submanifold $S$ and  {\it
 every} function $f$ the operator $\hat{X}_{S,f}$ is self-adjoint on
 $L^2(\Abar,\mu_{\rm AL})$ \cite{area}.
Thus the first conclusion  and Equation (\ref{div-X}) imply the
second conclusion of the theorem; i.e., that for every $\nu$ the
function ${F_{S,f}}^\nu\!_\nu$ is real valued.


\subsection{Proof of Lemma \ref{lm-al}}


Let us recall, that any regular, Borel, probability measure $\mu$
on the space $\Abar$ is uniquely determined by its projections $\mu_\gamma$
on the spaces $\Abar_\gamma$ (see Section (\ref{gen-conn})),
where we take into account all possible
analytic graphs $\gamma$ embedded in $\R^d$ \cite{al-meas}.
Hence, to prove the lemma it is enough to find out what
restrictions are imposed by the
assumptions of the lemma on each measure $\mu_\gamma$ .

Every $\mu_\gamma$ is a probability measure on the space
$\Abar_\gamma\cong G^N$, where $N$
is the number of edges of a given graph $\gamma$.
We outline now the properties of the group $G$ which will be used
below.

As every compact connected Lie group, $G$  is isomorphic \cite{woj} to a
quotient $\tilde{G}/M$,
where $M$ is a central discrete subgroup of $\tilde{G}$, and $\tilde{G}$ is a simple
product:
\[
\tilde{G}=T\times P
\]
of an abelian group\footnote{In fact, $T$ is isomorphic to a tori
$U(1)^n$ for some $n$.} $T$ and a semisimple group $P$ (obviously,
both $T$ and $P$ are compact and connected). The Lie algebras
$\tilde{G}'$ and $G'$ of  $\tilde{G}$ and $G$, respectively, are
isomorphic. The left (right) invariant vector fields on $G$ are
generated by curves:
\[
[-\epsilon,\epsilon]\ni s\mapsto g\exp(fs)\; (\exp(fs)g) \in G,
\]
where $\epsilon>0$, $f\in\tilde{G}'$, and $\exp(fs)\in\tilde{G}$.

We emphasize that, in what follows, we will regard $\exp(fs)$ as an
element of $\tilde{G}$ rather $G$.  Then, the
expression
\[
g\exp(fs)\in G
\]
is given by the (natural) action of $\exp(fs)\in\tilde{G}$ on $g\in G$.
The advantage is that every $\tilde{g}\in\tilde{G}$ can be expressed
as a product of elements of $T$ and $P$.


\subsubsection{Action of $\hat{X}_{C,f}$ on cylindrical functions compatible
with a given
graph}


We have assumed, that $\hat{X}_{C,f}$ is symmetric on $\cyl^\infty\subset
L^2(\Abar,\mu)$. Therefore,
\begin{equation}
0=\scal{1}{\hat{X}_{C,f}\Psi}=\int_{\Abar}\hat{X}_{C,f}\Psi \;d\mu,
\label{X-zero}
\end{equation}
for every $\Psi\in\cyl^{\infty}$.
We will fix a graph $\Gamma$  now, and investigate the consequences
of the equality (\ref{X-zero}), by using the cubes suitably
adjusted to the graph. The adjustment depends on the value
of the constant function $f$.

Let us consider an arbitrary graph $\gamma$ consisting of edges
$\{e_1,\ldots, e_N\}$. Divide each edge $e_I$ (by adding a new
vertex $v_I$) into edges $e_{I,i}$ ($i=1,2$) such that
$e_I=e_{I,1}\circ e_{I,2}$; a graph obtained in this way from
the graph $\gamma$ will be denoted by $\Gamma$. Notice that given
a measure $\mu$ on $\Abar$,  it is enough to consider only all the
graphs $\Gamma$ given by the construction above and by all the
graphs $\gamma$ in $\Sigma$, to reconstruct $\mu$ from the
measures $\mu_\Gamma$.

 Every cylindrical function $\Psi_\Gamma$ compatible with
$\Gamma$  can be written in the form:
\[
\Psi_\Gamma(\bar{A})=\psi(g_1(\bar{A}),g_2(\bar{A})),
\]
where:
\[
g_i(\bar{A}):=(\bar{A}(e_{1,i}),\ldots,\bar{A}(e_{N,i}))\in G^N,
\]
and $\psi$ is a function on $G^{2N}$.

Now, to draw the conclusions from Equation (\ref{X-zero}), we
consider two
separate cases: $(i)$ $f\in T'$, and $(ii)$ $f\in P'$, where $T'$ and $P'$
are the Lie
algebras of the
groups $T$ and $P$, respectively. We start with the more complicated case
of the semisimple algebra $P'$.

\setcounter{paragraph}{1}


\paragraph{\theparagraph\; The case of $f$ valued in $P'$}


Let $S$ consist of $N$ disjoint cubes $\{C_I\}$, such that each cube $C_I$
meets the graph
$\Gamma$ only at the vertex $v_I$ introduced above (as the division point
of an edge $e_I$ in the graph $\gamma$).
Let the orientation of each of the cubes coincide with the orientation
of the corresponding edge $e_I$. Suppose the function $f$ is defined
on $S$ in the following way:
\[
f|_{C_I}:={\rm const}_I=f_I\in P'
\]
(i.e. we {\em do not} assume, that $f_I=f_J$ for distinct $I,J$). Consider
the operator
\[
\hat{X}_{S,f}=\sum_{I=1}^{N}\hat{X}_{C_I,f_I}.
\]
Assuming that $\Psi_\Gamma$ is a smooth cylindrical function, using (\ref{X-zero})
we get
\begin{multline}
0=\int_{\Abar}\hat{X}_{S,f}\Psi_{\Gamma}\;d\mu=\int_{\Abar_\Gamma}
\hat{X}_{S,f}\psi\;d\mu_\Gamma=\\=-\frac{i}{2}\int_{G^{2N}}\frac{d}{ds}
\Big|_{s=0}\psi(g_1\exp(\vec{f}s),\exp(\vec{f}s)g_2)d\mu_\Gamma=\\=
-\frac{i}{2}\frac{d}{ds}\Big|_{s=0}\int_{G^{2N}}\psi(g_1\exp(\vec{f}s),
\exp(\vec{f}s)g_2)d\mu_\Gamma
\label{der}
\end{multline}
where
\begin{multline}
\vec{f}:=( f_1,\ldots, f_N)\in P^{\prime N}\ \ {\rm and}\ \\
\exp(\vec{f}s)=(\exp( f_1s),\ldots,\exp( f_Ns))\in P^N.
\label{expl}
\end{multline}
The two latter equations need some comments:
\begin{enumerate}
\item Notice that $f$ is valued in $\tilde{G}'=T'\oplus P'$, so in fact
expression
$\vec{f}\in P^{\prime N}$ means
\[
\vec{f}\in \{0_T\}\times P^{\prime N}
\]
where $0_T$ is zero in $T^{\prime N}$. Hence
\[
\exp(\vec{f}s)=(\mathbb{I}_{T},b),
\]
where $\mathbb{I}_{T}$ is the neutral element of $T^N$ and $b\in P^N$. Thus
 $\exp(\vec{f}s)\in\tilde{G}^N$ and the expression $g\exp(\vec{f}s)$ means a
natural action of
 an element of $\tilde{G}^N$ on $G^N=\tilde{G}^N/M^N$.
\item In the last step of Equation (\ref{der}) we have changed the order of
the integration on
 $G^{2N}$ and differentiation with respect to parameter $s$. To justify this we remark
  that we have integrated a continuously differentiable function over a
compact set
  $G^{2N}$ (see e.g. \cite{bill}).
\end{enumerate}
Equation (\ref{der}) holds for every smooth function on $G^{2N}$, in particular for
\[
\tilde{\psi}(g_1,g_2):=\psi(g_1\exp(\vec{f}s'),\exp(\vec{f}s')g_2).
\]
Setting $\tilde{\psi}$ into Equation (\ref{der}) we get
\[
0=\frac{d}{ds}\int_{G^{2N}}\psi(g_1\exp(\vec{f}s),\exp(\vec{f}s)g_2)\;d\mu_\Gamma,
\]
at an arbitrary value of $s$.
The fact that $\exp(\vec{f}s)$ is an element of the connected, compact group $P^N$
allows us to
write
\begin{equation}
\int_{G^{2N}}\psi(g_1b,bg_2)\;d\mu_\Gamma=\int_{G^{2N}}\psi(g_1,g_2)\;d\mu_\Gamma
\label{int-0}
\end{equation}
for every $b\in P^N$.

Now fix a function $\psi$ and define a map
\[
P^{N}\times P^N\ni(a,a')\mapsto\zeta(a,a'):=\int_{G^{2N}}\psi(g_1a,a'g_2)\;d\mu_\Gamma
\in \C.
\]
Notice that the function $\zeta$ is differentiable---indeed, it is defined as an
integral
of a smooth function over a compact set---hence to calculate a derivative of $\zeta$ we can
 first calculate an appropriate derivative of $\psi(g_1a,a'g_2)$ and then by integrating
 the result get the desired derivative of $\zeta$.

Equation (\ref{int-0}) implies immediately the following property of $\zeta$:
\begin{equation}
\zeta(ba,a'b)=\zeta(a,a')
\label{zeta}
\end{equation}
for every $b\in P^N$. Let $b=a^{-1}$. Then
\[
\zeta(a,a')=\zeta(\mathbb{I}_P,a'a^{-1})=:\xi(a'a^{-1}),
\]
where $\mathbb{I}_P$ is the neutral element of $P^N$. Now,  (\ref{zeta}) reads
\[
\xi(a'ba^{-1}b^{-1})=\xi(a'a^{-1}).
\]
The substitution $b_0=a'a^{-1}$ gives an identity
\[
\xi(b_0aba^{-1}b^{-1})=\xi(b_0),
\]
which holds for every $a,b,b_0\in P^N.$
Notice now, that if $L_1,L_2$ are arbitrary left invariant vector fields
on $P^N$, then the  vector $[L_1,L_2]_{b_0}$ tangent
to $P^N$ at the point $b_0$ can be generated by a curve of the form
\[
b_{0}a(t)b(t)a^{-1}(t)b^{-1}(t).
\]
Thus\footnote{Recall that $\zeta$ is differentiable, thus the function $\xi$ is
differentiable
 as well.}:
\[
[L_1,L_2]_{b_0} \xi=0
\]
But the group $P^N$ is semisimple, which means that
$[P^{\prime N},P^{\prime N}]=P^{\prime N}$.
Hence the function $\xi$  and consequently the function $\zeta$  both are
constant. This leads us to
 the following conclusion:
\begin{equation}
\int_{G^{2N}}\psi(g_1b_1,b_2g_2)\;d\mu_\Gamma=\int_{G^{2N}}\psi(g_1,g_2)\;d\mu_\Gamma
\label{int-semi}
\end{equation}
for every smooth function $\psi$ on $G^{2N}$ and for every $b_1,b_2\in P^{N}$.


\paragraph{\theparagraph\;The case of $f$ valued in $T'$}


Let $S$ consist now of $2N$ disjoint cubes $C_{I,i}$
($I=1,\ldots,N$ and $i=1,2$). Assume, that each $C_{I,i}$ meets
the graph $\Gamma$ in exactly  one point $y_{I,i}\in e_{I,i}$,
such that no $y_{I,i}$ is a vertex of $\Gamma$, and that the
orientation of each cube coincides with the orientation of the
corresponding edge. Let the function $f$ be
\[
f|_{C_{I,i}}={\rm const}_{I,i}=f_{I,i}\in T'.
\]
Then corresponding operator is
\[
\hat{X}_{S,f}=\sum_{i=1,2}\sum_{I=1}^{N}\hat{X}_{C_{I,i},f_{I,i}}.
\]
A derivation\footnote{In general the operator $\hat{X}_{S,f}$ is a sum of right and left
invariant vector fields on a product of some copies of $G$. But the abelian subgroup $T$
is a central subgroup of $T\times P=\tilde{G}$ and the elements of $T$ commute with all
the
 elements of $\tilde{G}$ (and of $G$). This means, that if $f$ is valued in $T'$, then
 corresponding left and right invariant vector fields defining the action of $\hat{X}_{S,f}$
 coincide and:
\[
\hat{X}_{S,f}\Psi_\Gamma=-i\frac{d}{ds}\Big|_{s=0}\psi(g_1\exp(\vec{f}_1s),
\exp(\vec{f}_2s)g_2),
\]
where $\vec{f}_i=( f_{1,i},\ldots, f_{N,i})\in T^{\prime N}$. Obviously we have
$\exp(\vec{f}_is)g_i=g_i\exp(\vec{f}_is)$.} similar to that one shown in the
previous paragraph  gives us the following counterpart of Equation (\ref{int-semi}):
\begin{equation}
\int_{G^{2N}}\psi(g_1t_1,t_2g_2)\;d\mu_\Gamma=\int_{G^{2N}}\psi(g_1,g_2)\;d\mu_\Gamma
\label{int-ab}
\end{equation}
for every $t_1,t_2\in T^N$.


\subsubsection{Final conclusion}


Let us now combine the results (\ref{int-semi}) and (\ref{int-ab}).
Because $T^N$ is a
 central subgroup of $T^N\times P^N$, every $t\in T^N$ commutes with every $b\in P^N$.
  This gives us:
\[
\int_{G^{2N}}\psi(g_1t_1b_1,t_2b_2g_2)\;d\mu_\Gamma=\int_{G^{2N}}\psi(g_1,g_2)\;d\mu_\Gamma
\]
for every $t_ib_i \in T^N\times P^N$, that is:
\begin{equation}
\int_{G^{2N}}\psi(g_1h_1,h_2g_2)\;d\mu_\Gamma=\int_{G^{2N}}\psi(g_1,g_2)\;d\mu_\Gamma
\label{res-0}
\end{equation}
for every $h_i\in\tilde{G}^N$ (or equivalently for every $h_i\in G^N$) and for every
smooth function $\psi$ on $G^{2N}$. By virtue of the Stone--Weierstrass theorem the closure
in the sup-norm of the  space $C^\infty(G^{2N},\C)$ is $C^0(G^{2N},\C)$, hence Equation
(\ref{res-0}) holds for every $\psi \in C^0(G^{2N},\C)$.

To show that $\mu_\Gamma$ is the Haar measure on $G^{2N}$ let us define the following
map on $G^{2N}$:
\[
(G^{N},G^N)\ni (g_1,g_2)\mapsto\omega(g_1,g_2):=(g_1,g^{-1}_2)\in(G^N,G^N).
\]
Then (\ref{res-0}) implies that the push forward  measure
\[
\mu_\Gamma^*:=\omega^*\mu_\Gamma
\]
is the right invariant Haar measure on $G^{2N}$. Indeed, a direct
calculation gives for every $(h_1,h_2)\in G^N\times G^N$
\[
\int_{G^{2N}}\psi(g_1,g_2)\;d\mu^*_\Gamma=\int_{G^{2N}}\psi(g_1h_1,g_2h^{-1}_2)\;
d\mu^*_\Gamma.
\]
Since $G$ is compact, the Haar measure $\mu^*_\Gamma$ is invariant
with respect to the map $\omega$, hence $\mu_\Gamma$ is the Haar
measure itself\footnote{The Haar measure $\mu^*_\Gamma$ on
$G^{2N}$ is a product of two copies of the Haar measure $\mu_{H}$
on $G^{N}$. On the other hand for every compact group left and
right invariant Haar measures on the group coincide \cite{raczka}.
Applying the latter fact to one of the $\mu_H$'s in the product
constituting $\mu^*_\Gamma$ we get the invariance of
$\mu^*_\Gamma$ with respect to $\omega$. Thus $\mu_\Gamma^*$ is
left and right invariant Haar measure on $G^{2N}$.}.

Every graph $\Gamma$ is obtained by a subdivision of some graph
$\gamma$, hence every cylindrical function $\Psi$ compatible with
$\gamma$ is also compatible\footnote{Notice that
$\Psi(\bar{A})=\psi(\bar{A}(e_I))=\psi(\bar{A}(e_{I,1})\bar{A}(e_{I,2}))$.}
with $\Gamma$. Thus making use of Equation (\ref{two-gra}) we
conclude, that the measure $\mu_\gamma$ on $\Abar_\gamma\cong G^N$
is the Haar measure. Because the graph $\gamma$ is arbitrary, 
\[
\mu=\mu_{\rm AL}.
\]
This completes the proof of lemma \ref{lm-al} and the proof of the
main theorem.
$\blacksquare$


\section{Summary}


In this paper we considered diffeomorphism covariant
$*$-representations of the Sahlmann holonomy-flux algebra
$\scripta$  (Definition \ref{def-sah}, Definition \ref{df-inv})
for a $G$-connection theory, where $G$ is an arbitrary compact
connected Lie group. We showed that if the principle bundle underlying the algebra is a trivial bundle\footnote{The result can be obtained in the general case of an arbitrary bundle $P(\Sigma,G)$, where $\Sigma$  is an arbitrary real-analytic manifold and $G$ is a compact
connected Lie group \cite{ol1}.} $\Sigma\times G$ and if $\Sigma=\R^d$, then the
carrier space used in Sahlmann's Characterization \ref{sahl-th} is
the orthogonal product (Theorem \ref{main}):
\begin{equation}
\h=\bigoplus_\nu L^2(\Abar,\mu_\nu),
\label{sum-hilb}
\end{equation}
where every measure $\mu_\nu$ is the natural measure:
\begin{equation}
\mu_\nu=\mu_{\rm AL}.
\label{nu-al}
\end{equation}
\bigskip

We emphasize that although the requirement of the diffeomorphism covariance of the representation singles out the measure $\mu_{\rm AL}$ there may possibly exist inequivalent diffeomorphism covariant representations of the Sahlmann algebra on the Hilbert space (\ref{sum-hilb}), which differ from each other in the family of functions $\{{F_{S,f}}^\iota\!_\nu\}$ (notice that the functions are not completely determined by Theorem \ref{main}). Thus the covariance with respect to the group of {\em analytic} diffeomorphisms does not seem to fix a unique diffemorphism covariant representation. However, an action of a larger group of diffeomorphisms (including certain non-analytic ones) is naturally defined on the algebra;  the covariance with respect to that group singles out precisely one (irreducible) representation of the Sahlmann algebra \cite{lost}.

An independent side result of our work is a proof showing that, if
a given measure $\mu$ on $\Abar$, the operators
$\{\hat{X}_{S,f}\}$ are symmetric in $\Cyl^\infty\subset
L^2(\Abar,\mu)$, for some sufficiently large class of submanifolds
$S\subset\Sigma$ and constant functions $f:S\rightarrow G'$, then
$\mu=\mu_{\rm AL}$ (Lemma \ref{lm-al}). This means that for every
manifold $\Sigma$ the $*$-representation
 $\pi$ of $\scripta$ defined on $\h= L^2(\Abar,\mu)$ as
 the identity i.e., the representation 
\[
\pi(\Phi)\Psi:=\Phi\Psi\;\;\;\;\pi(\hat{X}_{S,f})\Psi:=\hat{X}_{S,f}\Psi
\]
requires (\ref{nu-al}).
\bigskip

Another observation we made is that every $*$-representation of the
subalgebra $\cyl^\infty\subset\cyl$ can be uniquely extended to a
$*$-representation of the Ashtekar-Isham $C^\star$-algebra $\cyl$
(this fact enabled us to define the algebra $\scripta$ by means of
derivatives $\cal X$ and only {\em smooth} cylindrical functions).
\bigskip

{\bf Acknowledgements:} Our interest in this subject was stimulated
by discussions with Hanno Sahlmann who also explained to us many technical
points and discussed with us his unpublished considerations.
  Stanis{\l}aw L. Woronowicz and
 Jacek Tafel gave us very good hints helpful in proving Lemma \ref{pi-infty}
and Lemma \ref{lm-al}, respectively. Finally, we thank Abhay
Ashtekar and Thomas Thiemann for numerus discussions about the
quantum geometry and, respectively, QSD. This work was supported
in part by the following Polish KBN grants:
 2 PO3B 068 23,  2 PO3B 127 24.


\appendix



\section{Construction of the  diffeomorphism used in Le\-m\-ma \ref{diff-lm}}


In proving the main theorem we left lemma \ref{diff-lm} without any
proof. Let us now fill  that gap. The lemma states that:
\begin{lm}
For each graph $\gamma$  there exists an analytic diffeomorphism
$\varphi:\R^d\rightarrow\R^d$,
such that:
\begin{enumerate}
\item $\varphi$ preserves the cube $C_1$;
\item the graph  $\varphi(\gamma)$ has no edges transversal to the cube $C_2$.
\end{enumerate}
\end{lm}

{\bf Proof.} Let us recall definitions of cubes $C_i$ ($i=1,2$):
\begin{gather*}
C_1:=\{y\in C \; \big| \; x^1(y)\leq 0\}\\
C_2:=\{y\in C \; \big| \; x^1 (y)> 0\},
\end{gather*}
where
\[
C:=\{y\in \R^d\;\big|\;
\begin{cases}
x^d(y)=0&\\
-l< x^1(y) \leq l &\\
-l< x^i(y) < l & \text{for $i=2,3,\ldots,d-1$.}
\end{cases};
\;\;\;l>0\}.
\]
Here $(x^j)$, ($j=1,\ldots,d$) is a coordinate frame on $\R^d$ obtained by an affine
transformation of the Cartesian coordinates on $\R^d$.

Assume that the set $\{\tilde{e}_{J}\}$ contains all edges of the graph $\gamma$ such that
each $\tilde{e}_{J}$ intersects the cube $C_2$ in isolated points. Because the set
$\{\tilde{e}_J\}$ is finite and the edges and cube $C_2$ are analytic, the set
$W:=(\bigcup_{J}\tilde{e}_{J})\cap C_2\subset \R^d$ (i.e. the set of intersection
points of the cube and the edges) is finite as well. Define
\[
\epsilon:={\rm min}\{x^1(y)\;\big|\;y\in W\}.
\]
Clearly $\epsilon>0$. Let us assume now that the function $\varphi':\R\rightarrow\R$ is
an analytic diffeomorphism such that:
\begin{equation}
\begin{cases}
\varphi'(-l)=-l&\\
\varphi'(0)=0&\\
\varphi'(\epsilon)>l
\end{cases}.
\label{varphi}
\end{equation}
Then we can define
\[
\varphi^i(x^j):=
\begin{cases}
\varphi'(x^1)&\text{for $i=1$},\\
x^i&\text{otherwise}.
\end{cases}
\]
In this way we have reduced our task to constructing the diffeomorphism $\varphi'$.

To construct the diffeomorphism it will be convenient to transform the coordinate $x^1$  in the following way:
\[
\R\ni x^1\mapsto z(x^1):=\exp[\frac{\ln 2}{l}x^1]\in ]0,\infty[.
\]
Values of the new coordinate at points relevant for us  are the following:
\begin{gather*}
z(-l)=\frac{1}{2},\;\;\;z(l)=2,\\
z(0)=1,\;\;\;z(\epsilon)=1+\epsilon'
\end{gather*}
for some $\epsilon'>0$. Consider now two families of analytic diffeomorphisms on
$]0,\infty[$
\begin{gather*}
\alpha_t(z):=\frac{(1+t)^z-1}{2t}\\
\beta_s(z):=\frac{\ln(1+sz)}{2\ln(1+s)},
\end{gather*}
where $t,s\in]0,\infty[$. We have
\begin{gather}
\lim_{t\rightarrow\infty}\alpha_t(z)=
\begin{cases}
\infty&\text{for $z>1$}\\
\frac{1}{2}&\text{for $z=1$}\\
0&\text{for $z<1$}
\end{cases}\label{alpha}\\
\lim_{s\rightarrow\infty}\beta_s(z)=\frac{1}{2};\;\;\;
\lim_{s\rightarrow 0}\beta_s(z)=\frac{1}{2}z\label{beta}.
\end{gather}
Equation (\ref{alpha}) means that it is possible to find $t_0$, such that:
\[
\alpha_{t_0}(1+\epsilon')>2
\]
and:
\[
\alpha_{t_0}(\frac{1}{2})<\frac{1}{4}.
\]
From (\ref{beta}) we know that for some $s_0$
\[
\beta_{s_0}(\frac{1}{2})=\frac{1}{2}-\alpha_{t_0}(\frac{1}{2}).
\]
Then
\[
\phi(z):=\alpha_{t_0}(z)+\beta_{s_0}(z)
\]
is an analytic diffeomorphism on $]0,\infty[$ such that:
\[
\begin{cases}
\phi(\frac{1}{2})=\frac{1}{2}&\\
\phi(1)=1&\\
\phi(1+\epsilon')>2
\end{cases},
\]
which coincides with (\ref{varphi}). Thus
\[
\varphi'(x^1):=\frac{l}{\ln2}\ln[\phi(z(x^1))].
\]
$\blacksquare$



\begin{thebibliography}{}
\bibitem{sahl-1} Sahlmann H 2002 Some Comments on Representation Theory of
the Algebra Underlying Loop Quantum Gravity {\it Preprint} gr-qc/0207111
%
\bibitem{sahl-2} Sahlmann H 2002 When Do Measures on the Space of Connections
Support the Triad Operators of Loop Quantum Gravity? {\it Preprint} gr-qc/0207112
%
\bibitem{sahlthiem} Sahlmann H and Thiemann T 2003 On the Superselection Theory of the Weyl Algebra for Diffeomorphism Invariant Quantum Gauge Theories {\it Preprint} gr-qc/0302090
%
\bibitem{al-difgeom} Ashtekar A and Lewandowski J 1995 Differential
geometry on the space of connections using projective techniques,
{\it Jour. Geo. \& \ Phys.} \textbf{17}, 191--230
%
\bibitem{geomgen}
Rovelli C and Smolin L 1995 Discreteness of area and volume
in quantum gravity \textit{Nucl. Phys.} \textbf{B442} 593--622 
Erratum: \textit{Nucl. Phys.} \textbf{B456} 753 {\it Preprint} gr-qc/9411005 \\
%
Loll R 1996 The volume operator in discretized quantum gravity
\textit{Phys. Rev. Lett.} \textbf{75} 3048--3051 {\it Preprint} gr-qc/9506014 \\
%
Thiemann T 1998 Closed formula for the matrix elements of the volume operator in canonical quantum gravity \textit{J. Math. Phys.} \textbf{39} 3347-3371 {\it Preprint} gr-qc/9606091
%
\bibitem{lenght} Thiemann T 1998 A length operator for canonical quantum gravity {\it J. Math. Phys.} {\bf 39} 3372--3392 {\it Preprint} gr-qc/9606092
%
\bibitem{area} Ashtekar A and Lewandowski J 1997 Quantum theory of geometry.
I: Area operators {\it Class. Quant. Grav.} {\bf 14} A55--A82 {\it Preprint} gr-qc/9602046
%
\bibitem{vol} Ashtekar A and  Lewandowski J 1998 Quantum Theory of Geometry. 2. Volume operators {\it Adv. Theor. Math. Phys.} {\bf 1} 388--429 {\it Preprint} gr-qc/9711031
%
\bibitem{acz} Ashtekar A, Corichi A and  Zapata J A 1998 Quantum Theory of Geometry III: Non-commutativity of Riemannian Structures {\it Class. Quant. Grav.} {\bf 15} 2955--2972 {\it Preprint} gr-qc/9806041
%
\bibitem{lqg}
Rovelli C and Smolin L 1990 Loop representation for quantum general relativity \textit{Nucl. Phys. } \textbf{B331} 80--152 \\
%
Rovelli C and Smolin L 1994 The physical Hamiltonian in nonperturbative quantum gravity \textit{Phys. Rev. Lett.} \textbf{72} 446--449 {\it Preprint} gr-qc/9308002 \\
%
Ashtekar A, Lewandowski J, Marolf D, Mour{\~a}o J and  Thiemann T 1995 Quantization of diffeomorphism invariant theories of connections with local degrees of freedom {\it J. Math. Phys.} {\bf 36}  6456--6493, {\it Preprint} gr-qc/9504018 \\
%
Rovelli C 1998  Loop Quantum Gravity  {\it Living Rev. Rel.} {\bf 1} 1 {\it Preprint} gr-qc/9710008 
%
\bibitem{QSD} Thiemann T 1996 Anomaly-free formulation of non-perturbative, four-dimensional Lorentzian quantum gravity \textit{Phys. Lett.} \textbf{B380} 257--264 {\it Preprint} gr-qc/9606088\\
%
Thiemann T 1998 Quantum Spin Dynamics (QSD) \textit{Class. Quant. Grav.}  \textbf{15} 839--873 {\it Preprint} gr-qc/9606089\\
%
Thiemann T 1998 QSD III : Quantum Constraint Algebra and Physical Scalar Product in Quantum General Relativity \textit{Class. Quant. Grav.} \textbf{15} 1207--1247 {\it Preprint} gr-qc/9705017 \\
%
Thiemann T 1998 QSD V : Quantum Gravity as the Natural Regulator of Matter Quantum Field Theories \textit{Class. Quant. Grav.} \textbf{15} 1281--1314 gr-qc/9705019
%
\bibitem{al-meas} Ashtekar A and  Lewandowski J 1994 Representation theory of analytic holonomy $C^\star$-algebras {\it Knots and quantum gravity} (Baez J (ed), Oxford: Oxford University Press) {\it Preprint} gr-qc/9311010 \\
%
Marolf D and Mour\~ao J 1995 On the support of the Ashtekar-Le\-wan\-do\-wski measure \textit{Commun. Math. Phys.} \textbf{170}, 583--606. \\
%
Ashtekar A and Lewandowski J  1995 Projective techniques and functional integration for gauge theories {\it J. Math. Phys.} {\bf 36} 2170--2191 {\it Preprint} gr-qc/9411046
%
\bibitem{B} Baez J C  1994  Generalized measures in gauge theory \textit{Lett. Math. Phys.} \textbf{31} 213--223 {\it Preprint} hep-th/9310201 
%
\bibitem{ai} Ashtekar A and Isham C J 1992 Representations of the holonomy algebras for gravity and non-Abelian gauge theories {\it Class. Quant. Grav.} {\bf 9} 1433--1467
%
\bibitem{smooth} 
Lewandowski J and Marolf D  Loop constraints: A habitat and their algebra, \textit{Int. J. Mod. Phys.} \textbf{D7} 299--330 {\it Preprint} gr-qc/9710016 \\
%
Baez J C and Sawin S 1997 Functional integration on spaces of connections \textit{Jour. Funct. Analysis} \textbf{150} 1--27 {\it Preprint} q-alg/9507023 \\
%
Lewandowski J and Thiemann T 1999 Diffeomorphism invariant quantum field theories of connections in terms of webs \textit{Class. Quant. Grav.} \textbf{16} 2299--2322 {\it Preprint} gr-qc/9901015
%
%
\bibitem{woj} Wojty{\'n}ski W 1986 {\it Grupy i algebry Liego (Lie groups and algebras)}
(Warszawa: PWN --- Polish Scientific Publisher)
%
\bibitem{bill} Billingsley P 1979 {\it Probability and measure} (New York-Chichester-Brisbane-Toronto: John Wiley \& Sons)
%
\bibitem{raczka} Barut A O and  R\c{a}czka R 1977 {\it Theory of group representations and applications} (Warszawa: PWN --- Polish Scientific Publisher)

\bibitem{ol1} Oko{\l}\'ow A  and Lewandowski L Authomorphism covariant representations of the holonomy-flux $*$-algebra, in preparation

\bibitem{lost} Lewandowski J, Oko{\l}\'ow A, Sahlmann H and Thiemann T, in preparation
\end{thebibliography}
\end{document}